\documentclass[%
 superscriptaddress,
 amsmath,amssymb,
 aps,
twocolumn,
english
]{revtex4-2}

\usepackage{graphicx}
\usepackage{dcolumn}
\usepackage{bm}
\usepackage[dvipsnames]{xcolor}
\usepackage{enumerate}
\usepackage{braket}

\begin{document}


\title{Superradiance and subradiance in inverted atomic arrays}

\author{Oriol Rubies-Bigorda}
\email{orubies@mit.edu}
\affiliation{Department of Physics, Massachusetts Institute of Technology, Cambridge, Massachusetts 02139, USA}
\affiliation{Department of Physics, Harvard University, Cambridge, Massachusetts 02138, USA}
\author{Susanne F. Yelin}
\affiliation{Department of Physics, Harvard University, Cambridge, Massachusetts 02138, USA}

\begin{abstract}
Superradiance and subradiance are collective effects that emerge from coherent interactions between quantum emitters. Due to their many-body nature, theoretical studies of extended samples with length larger than the atomic transition wavelength are usually restricted to their early time behavior or to the few-excitation limit. We use herein a mean-field approach to reduce the complex many-body system to an effective two-atom master equation that includes all correlations up to second order and that can be numerically propagated in time. We find that three-dimensional and two-dimensional inverted atomic arrays sustain superradiance below a critical lattice spacing and quantify the scaling of the superradiant peak for both dimensionalities. Finally, we study the late-time dynamics of the system and demonstrate that a subradiant phase appears before the system finally relaxes.
\end{abstract}

\maketitle

\section{Introduction}

\begin{figure}
\includegraphics[width=\linewidth]{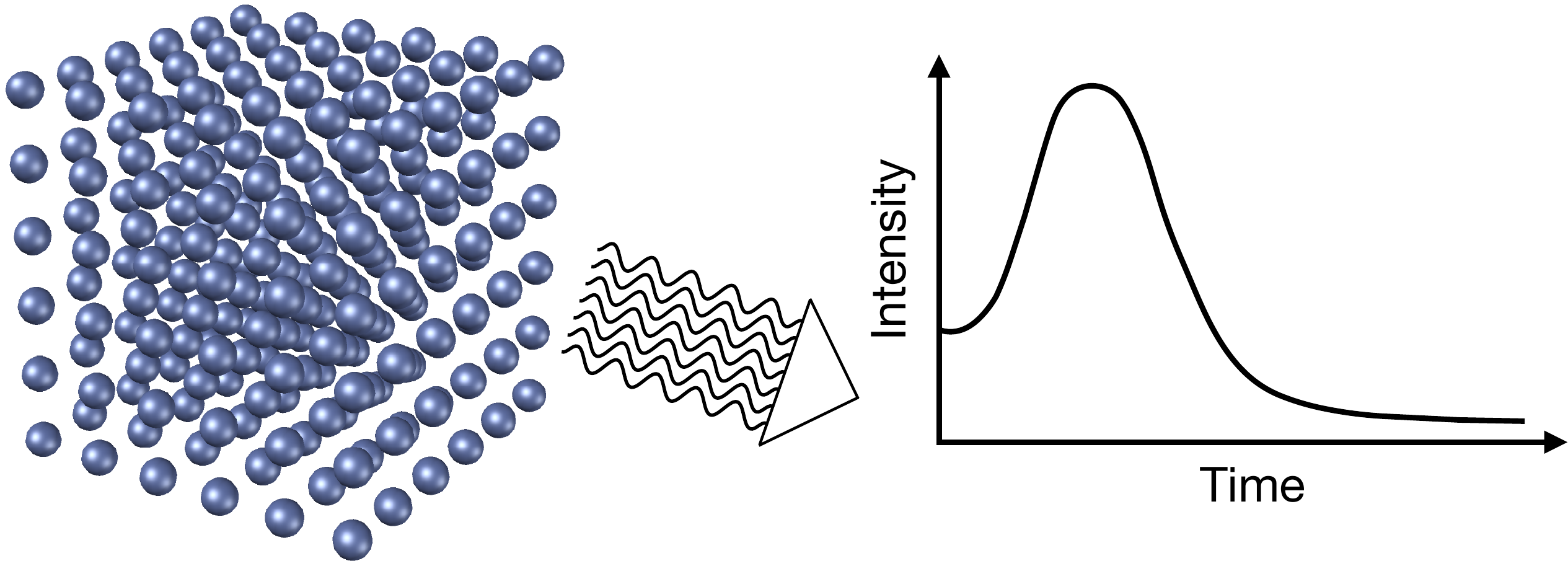}
\caption{\label{fig: general_scheme_superradiace} \textbf{Superradiance in atomic arrays} Collective emission of light from a three-dimensional array of closely spaced  dipole-coupled atomic emitters. The radiated intensity grows at early times, giving rise to the superradiant burst.}
\end{figure}

The decay and interaction of dilute ensembles of two-level atoms with the radiation field is commonly described by the semiclassical Maxwell-Bloch equations, which assume the atoms to emit independently and result in an exponential decay of the atomic excitation. While this approximation is accurate when emitters are separated by large distances, it breaks down for dense media. As first noted by Dicke, the photon emitted by one atom can coherently interact with close-by atoms and therefore stimulate emission of additional photons \cite{intro_Dicke,intro_Lehmberg,intro_gross}. As a result, the atomic dipoles lock in phase, build up coherences and collectively emit at a higher rate, giving rise to the \textit{superradiant burst} in Fig.~\ref{fig: general_scheme_superradiace}. This phenomenon has been experimentally observed in a wide variety of systems, ranging from thermal gases \cite{Experiment_1,Experiment_2,Experiment_thermal} to Bose-Einstein condensates \cite{Experiment_3,Bose_cond_2} and Rydberg atoms \cite{Experiment_4,Rydberg_2,Experiment_Rydberg}.

In the simplest model, one assumes all atoms to lie in the same spatial position, such that they cannot be distinguished. Then, the $N$-atom system can only be in one of the $N+1$ symmetric states and the maximum intensity of the emitted light pulse for an initially inverted sample scales with $N^2$ \cite{intro_Dicke,intro_gross}, as opposed to the linear scaling characteristic of independent emitters. In extended samples larger than one atomic transition wavelength, dipole-dipole interactions between different atoms become relevant and the aforementioned symmetry is broken. As a result, the whole Hilbert space with dimension $2^N$ needs to be considered and most theoretical studies of superradiance and subradiance are consequently restricted to the emission of few photons \cite{single_excitation_1,single_excitation_2,single_excitation_3,single_excitation_4,single_excitatin_5,single_excitation_6} or to systems with a small number of atoms \cite{few_atoms_charmichael,few_atoms_Anna,charmichael_2} such that numerical Monte Carlo-type methods can be applied. 

The recent experimental advances in optical lattices \cite{OpticalLAttice_1,OpticalLAttice_2,OpticalLAttice_3} and atomic tweezers \cite{Tweezer_1,Tweezer_2,Tweezer_3}, which allow one to produce atomic lattices (as well as more complex configurations) with interparticle spacing of the order of an atomic transition wavelength, have revived the interest in superradiant and subradiant physics. While these systems have been extensively studied in the case where only one excitation is present in the lattice \cite{OneExcitation_Ephi,OneExcitation_Ana, OneExcitation_Chang, one_exc_array_mariona,one_exc_array_taylor,one_exc_array_adams,one_excitation_roustekoski,OneExcitation_me}, the behavior of inverted lattices is poorly understood. Recent theoretical studies have shown that the appearance of the superradiant burst in inverted samples is determined by the statistics of the first two photons \cite{superradiantburst_ana,Ana_super_new}, or alternatively by the Taylor series expansion of the photon emission rate at time $t=0$ \cite{robicheaux_superradiance}. While these methods allow one to determine the superradiant phase diagram and the initial slope of the emitted radiation, they provide no information about the scaling of the superradiant peak or the nature of the subsequent time evolution.

In this work, we use an alternative method developed in Refs.~\cite{Fleishbauer,AMO_GuindarLin,Hanzhen} based on a mean-field approach that includes all correlations up to second order. By tracing out the degrees of freedom of $N-2$ atoms and the radiation field, one can reduce the description of the full many-body system to an effective nonlinear two-atom master equation, which can be numerically propagated in time. We herein confirm the appearance of a superradiant burst
in two-dimensional and three-dimensional atomic arrays with small enough interparticle spacing and extend the results in Refs.~\cite{superradiantburst_ana, robicheaux_superradiance,Ana_super_new} by characterizing the scaling of the superradiant peak for both dimensionalities, as well as by studying the long-time dynamics of the system, which exhibits a subradiant behavior.

\section{Formalism}

\begin{figure}[b]
\includegraphics[width=\linewidth]{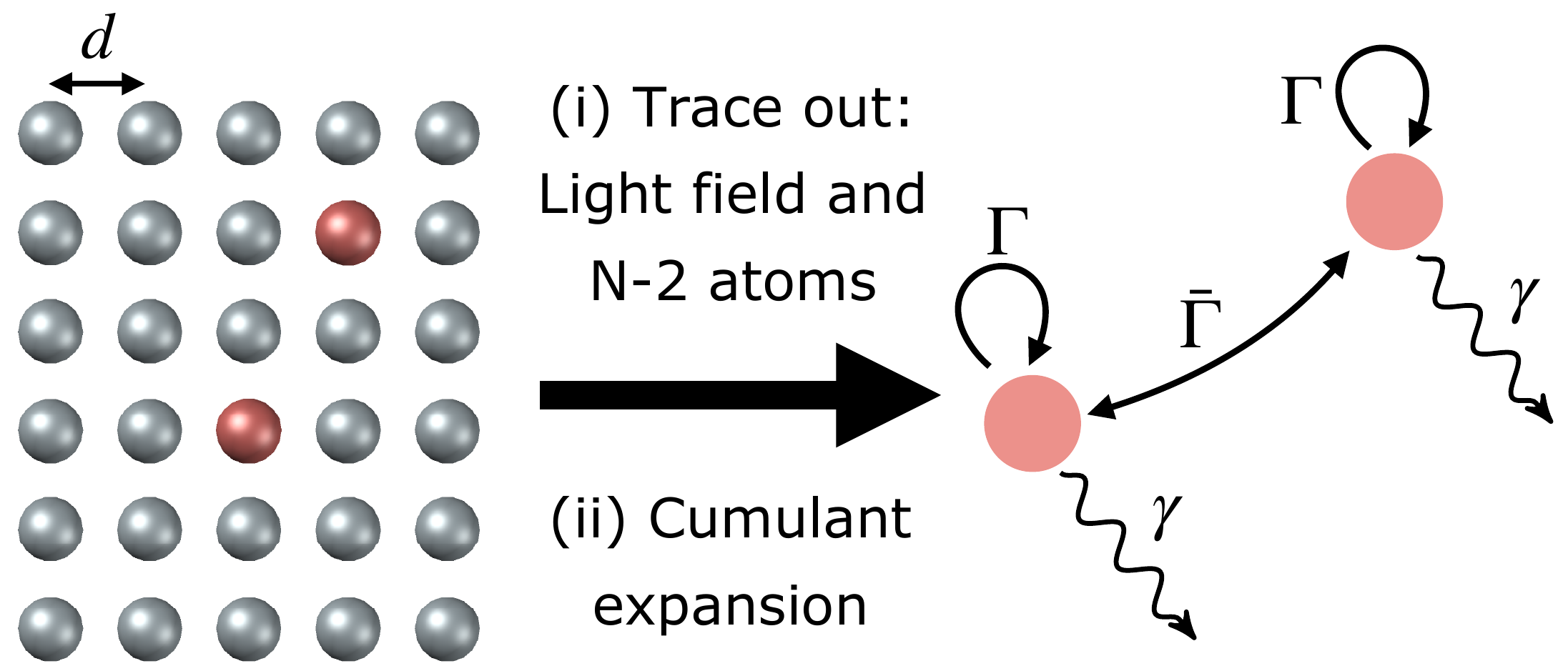}
\caption{\label{fig: sketch_MasterEquation} \textbf{Reduced two-atom system} Two probe atoms, represented in red, are chosen from an $N$-atom array with interparticle spacing $d$. Tracing out the degrees of freedom from the radiation field and the $N-2$ nonselected atoms and performing a cumulant expansion results in a master equation for the reduced two-atom system. The evolution of the system depends on three decay rates that arise from the dipole-dipole interactions between all array atoms mediated by the electromagnetic field: the spontaneous decay rate $\gamma$ and the cooperative single-particle and two-particle decay rates $\Gamma$ and $\bar{\Gamma}$ respectively. }
\end{figure}

We first summarize the formalism derived in full detail in Ref.~\cite{Hanzhen} that reduces the description of the atomic array to a two-atom master equation \footnote{We refer the reader to Ref.~\cite{Hanzhen} for a thorough derivation of the full formalism described in the main text.}. We consider an ensemble of $N$ two-level atoms that interact with the vacuum electromagnetic field. The Hamiltonian of the system can be written as the sum of three terms: the free Hamiltonian of the atoms $H_\textit{atoms}$, the free Hamiltonian of the field $H_\textit{field}$, and the interaction term in the dipole approximation $H_\textit{int}=-\sum_i \vec{p}_i \vec{E}(\vec{r}_i)$, where the index $i$ labels the lattice atoms, $\vec{p}$ is the dipole operator and $\vec{E}(\vec{r}_i)$ represents the quantized field at the atomic positions. Then, two probe atoms labeled as $i \in \{ 1,2 \}$ are selected (as illustrated by the two red particles in Fig.~\ref{fig: sketch_MasterEquation}) and the Hamiltonian is split into two parts $H=H_0+V$ such that $V$ contains the interaction of the two atoms with the field and $H_0$ includes the rest of the terms
\begin{align}
    H_0 & =H_\textit{atoms}+H_\textit{field}-\sum_{i \neq 1,2} \vec{p}_i \vec{E}(\vec{r}_i), \nonumber \\
    V & =-\sum_{i=1,2} \vec{p}_i \vec{E}(\vec{r}_i).
\end{align}

Moving to the interaction picture and tracing over the environment degrees of freedom, that is, the radiation field and the $N-2$ nonselected atoms, one can obtain the effective time evolution operator of the reduced system on the Schwinger-Keldysh contour \cite{keldysh}. Using the Markov and the rotating-wave approximations and performing a cumulant expansion that keeps all correlations up to second order finally results in a master equation for the two probe atoms. 
We additionally consider randomly polarized two-level atoms and neglect retardation effects of the electromagnetic field such that all changes in the atomic variables propagate instantaneously. In that case, the coordinate dependence of the reduced density matrix for the two-probe atom can be dropped \cite{AMO_GuindarLin,Hanzhen}. The dynamics of the reduced system are then described by the quantities
\begin{align}
\label{eq: relevant_variables}
    a & = \langle \frac{\hat{\sigma}_{ee}^{(1)} + \hat{\sigma}_{ee}^{(2)}}{2} \rangle = \rho_{ee,ee}+\frac{\rho_{ee,gg}+\rho_{gg,ee}}{2}, \nonumber \\
    n & = \langle \hat{\sigma}_z^{(1)} \hat{\sigma}_z^{(2)} \rangle = \rho_{ee,ee}-\rho_{ee,gg}-\rho_{gg,ee}+\rho_{gg,gg}, \nonumber \\
    \rho_{eg,ge} & = \langle \hat{\sigma}_-^{(1)} \hat{\sigma}_+^{(2)} \rangle,
\end{align}
where we have defined the density-matrix elements $\rho_{\alpha\beta,\gamma\delta}=\langle \alpha_1 \gamma_2 | \rho | \beta_1 \delta_2 \rangle$ as well as the operators $\hat{\sigma}_{ee}^{(i)}=\ket{e^{(i)}} \bra{e^{(i)}}$, $\hat{\sigma}_{+}^{(i)}=\ket{e^{(i)}} \bra{g^{(i)}}$, $\hat{\sigma}_{-}^{(i)}=\ket{g^{(i)}} \bra{e^{(i)}}$, and $\hat{\sigma}_{z}^{(i)}=\ket{e^{(i)}} \bra{e^{(i)}}-\ket{g^{(i)}} \bra{g^{(i)}}$. These three variables have a clear physical meaning. The $a$ represents the average upper-level population in the system, such that $-\dot{a}$ directly gives the emitted intensity per particle. The $n$ is the average value of the spin-spin correlation $\hat{\sigma}_z^{(1)} \hat{\sigma}_z^{(2)}$, which takes a maximum value ($n=1$) when both atoms are either excited or deexcited and a minimum value ($n=-1$) when only one of the atoms is in the excited state. Together with $a$, it fully determines the populations of both probe atoms and we therefore refer to it as the effective two-atom inversion of the system. Finally, $\rho_{eg,ge}$ corresponds to the two-atom flip–flop term $ \langle \hat{\sigma}_-^{(1)} \hat{\sigma}_+^{(2)} \rangle$ and quantifies the coherence built between the probe atoms, that is, between the single-excitation states $\ket{e^{(1)},g^{(2)}}$ and $\ket{g^{(1)},e^{(2)}}$.

The resulting equations of motion can be written as
\begin{align}
\label{eq: equationsmotion}
    \dot{a} & = -(2\Gamma+\gamma) a + \Gamma, \nonumber \\
    \dot{n} & = -2(2\Gamma+\gamma) n -2\gamma(2a-1) + 8\bar{\Gamma} \rho_{eg,ge}, \nonumber \\
    \dot{\rho}_{eg,ge} & = -(2\Gamma+\gamma) \rho_{eg,ge} + \bar{\Gamma} n 
\end{align}
and depend on three decay rates, as depicted in Fig.~\ref{fig: sketch_MasterEquation}. The first is the spontaneous decay rate of a single atom in the presence of the vacuum field $\gamma$. The second and the third are the cooperative decay rates $\Gamma$ and $\bar{\Gamma}$, which arise from the interaction with the remaining $N-2$ atoms mediated by the electromagnetic field. Here $\Gamma$ can be understood as a self-energy or self-decay rate that comes into the reduced master equation through terms involving raising and lowering operators of one probe atom only (e.g. $\hat{\sigma}_-^{(1)} \hat{\sigma}_+^{(1)}$). As for the two-atom decay rate $\bar{\Gamma}$, it describes the effective interaction between both probe atoms (see the sketch in Fig.~\ref{fig: sketch_MasterEquation}) and appears through cross terms such as $\hat{\sigma}_-^{(1)} \hat{\sigma}_+^{(2)}$. Note that, additionally, the interaction between the atoms generates a cooperative energy shift. Because such shifts are generally small in two-level atoms \cite{shifts_small_gray2016}, here we set it to zero.

As shown in Ref.~\cite{Hanzhen}, a closed form can be found for the collective decay rates
\begin{widetext}
\begin{align}
\label{eq: collectivedecayrates_main}
\Gamma(\vec{r}) &= \frac{\wp^4}{\hbar^4} \frac{2a}{\gamma/2+\Gamma} \sum_{\vec{x}}  \left| \tilde{D}^{ret} (\vec{r}-\vec{x}) \right|^2 + \frac{\wp^4}{\hbar^4} \frac{2\rho_{eg,ge}}{\gamma/2+\Gamma} \sum_{\vec{x}_1} \sum_{\vec{x}_2} \tilde{D}^{ret} (\vec{r}-\vec{x}_1) \tilde{D}^{*ret} (\vec{r}-\vec{x}_2), \nonumber \\
\bar{\Gamma}(\vec{r}_1,\vec{r}_2) &=  \frac{\wp^4}{\hbar^4} \frac{2a}{\gamma/2+\Gamma} \sum_{\vec{x}} \tilde{D}^{ret} (\vec{r}_1-\vec{x}) \tilde{D}^{*ret} (\vec{r}_2-\vec{x}) + \frac{\wp^4}{\hbar^4} \frac{2\rho_{eg,ge}}{\gamma/2+\Gamma} \sum_{\vec{x}_1} \sum_{\vec{x}_2}  \tilde{D}^{ret} (\vec{r}_1-\vec{x}_1) \tilde{D}^{*ret} (\vec{r}_2-\vec{x}_2),
\end{align}
\end{widetext}
where $\wp$ is the dipole matrix element. These collective decay rates involve summations over all lattice atoms, located at positions $\vec{x}$, and consequently depend on the size or number of particles of the system. Here $\Gamma$ and $\bar{\Gamma}$ additionally depend on the state of the atomic system, characterized by the variables $a$, $n$, and $\rho_{eg,ge}$, and therefore vary over time during the decay process. 
If all atoms are in the ground state ($a=\rho_{eg,ge}=0$ and $n=1$), both collective decay rates are zero and the equations of motion given in Eq.~(\ref{eq: equationsmotion}) reduce to $\dot{a}=\dot{n}=\dot{\rho}_{eg,ge}=0$. That is, the ground state of the system is simultaneously its steady state, as expected in the absence of an external driving field. 
Importantly, $\Gamma$ and $\bar{\Gamma}$ also depend on the specific positions of the probe atoms, $\vec{r}_1$ and $\vec{r}_2$. 
This dependence, however, is much weaker than that of the retarded Green's function in the medium $\tilde{D}^{ret}$, as it is washed out by the summation over all lattice atoms located at positions $\vec{x}$. 
To account for it and to obtain the behavior representative of the whole atomic ensemble, we consider and compare two different ways of computing the two-atom cooperative decay rate. The first, labeled as $\bar{\Gamma}^{(n.n.)}$, assumes that the two probe atoms are nearest neighbors, while the second, labeled as $\bar{\Gamma}^{(av.)}$, considers an average over different positions of the atom pairs (refer to Appendix~\ref{appendix: decayrates} for a more detailed discussion).

Finally, the collective decay rates, and therefore the evolution of the system, depend on the retarded Green's function in the medium $\tilde{D}^{ret}$. This quantity describes the propagation of the electromagnetic field in the presence of the ensemble of atoms and therefore depends on the dimension of the system. It can be obtained from the free-space Green's function and the polarization of the medium by means of the Dyson equation formalism \cite{Dyson}, as described in Appendix~\ref{appendix: greensFunction}. For a three-dimensional ensemble of randomly polarized atoms, it can be written as
\begin{align}
\label{eq: retardedGreen_3D}
    \tilde{D}_{3D}^{\textit{ret}}(r) & =-\frac{i\hbar k_0^2}{6 \pi \epsilon_0}\frac{e^{-i k_0 r} e^{\xi r}}{r} , \nonumber \\
    \xi &= \gamma \frac{2a-1}{\gamma/2+\Gamma} \frac{\pi}{k_0^2 d^3} ,
\end{align}
where $d$ is the lattice spacing, $k_0=2\pi/\lambda$ is the wave number associated with the atomic transition wavelength $\lambda$, and $r=\sqrt{x^2+y^2+z^2}$. In a two-dimensional medium such that all atoms are at $z=0$, it takes the form
\begin{align}
\label{eq: retardedGreen_2D}
    \tilde{D}_{2D}^{ret}(\rho)
       & =-\frac{i \hbar k_0}{6 \pi \epsilon_0}  \int_{0}^{\infty} dq \frac{q J_0(q\rho)}{\sqrt{ q^2/k_0^2-1+2i\epsilon/k_0}-i \chi} ,\nonumber \\
       \chi & =\gamma \frac{2a-1}{\gamma/2+\Gamma} \frac{\pi}{k_0^2 d^2} ,
\end{align}
where $\rho=\sqrt{x^2+y^2}$ is the distance on the plane defined by the two-dimensional array and the small positive constant $\epsilon$ ensures the convergence of the integral.

Both Green's functions are complex valued, oscillate with distance, and their absolute values are increasing functions of the upper-level population $a$.
While the two-dimensional Green's function in the medium $\tilde{D}_{2D}^{ret}(\rho)$ decreases with distance $\rho$ for all values of $a$, its three-dimensional counterpart $\tilde{D}_{3D}^{ret}(r)$ increases with distance $r$ for $a>\frac{1}{2}$. In that case, the three-dimensional atomic array turns into an amplifying medium.

Equations (\ref{eq: equationsmotion}) and (\ref{eq: collectivedecayrates_main}), together with the expressions of the retarded Green's functions in the medium, form a self-consistent set of equations that can be numerically propagated in time to obtain the dynamics of the system.

\begin{figure}
\includegraphics[width=\linewidth]{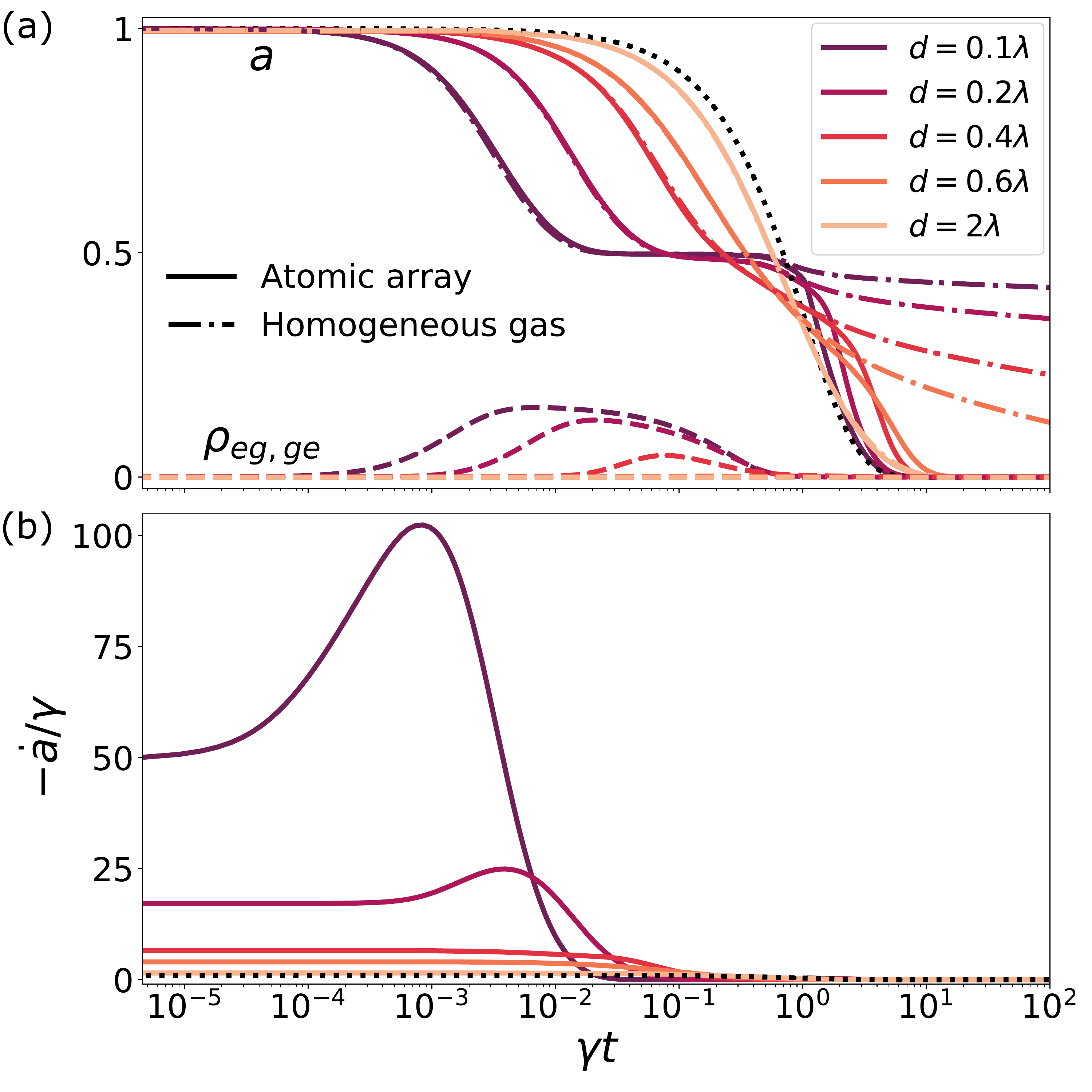}
\caption{\label{fig: time_evolution} \textbf{Time evolution of three-dimensional atomic arrays} (a) Average upper-level population $a$ (solid lines) and two-atom coherence $\rho_{eg,ge}$ (dashed lines) as a function of time for a spherical three-dimensional atomic array with $N_{rad}=25$ particles in the radial direction (and  $7153$ atoms in total) and for different interparticle spacings $d$. The black curve represents the decay in the absence of interactions between particles, that is, it recovers the limit $d/\lambda \rightarrow \infty$. The dash-dotted traces represent the dynamics for a homogeneous spherical gas of atoms with the same radius and atomic density. The two-atom coherence is nearly identical in both cases. (b) Intensity per particle $-\dot{a}$ as a function of time for atomic arrays with different spacings. The color code is the same as in (a). These results are obtained using the averaged collective decay rate $\bar{\Gamma}^{(av.)}$.}
\end{figure}

\section{Three-dimensional atomic arrays}

Figure~\ref{fig: time_evolution}(a) shows the resulting dynamics for a spherical, three-dimensional atomic array with $N_{rad}=25$ particles in the radial direction and obtained with the averaged decay rate $\bar{\Gamma}^{(av.)}$. As shown in Appendix~\ref{appendix: 3D array}, the major features for three-dimensional lattices are independent of the specific form considered for the two-atom cooperative decay rate. 
For a small interparticle spacing of $d=0.1\lambda$, the average upper-level population $a$ (purple solid leftmost curve) initially decays at a much faster rate than would occur for non-interacting particles (black dotted curve). The decay rate or emitted intensity per particle $-\dot{a}$, given by the purple (upper) trace in Fig.~\ref{fig: time_evolution}(b), increases at early times and a superradiant burst appears. This substantial increase of emission results from the buildup of coherences in the system, as illustrated by the two-particle coherence $\rho_{eg,ge}$ [purple dashed curve in Fig.~\ref{fig: time_evolution}(a)]. After the initial superradiant decay, a subradiant phase appears. The emitted intensity is heavily suppressed and the average upper-level population remains roughly constant, while the coherences built up during the superradiant burst slowly decay. As soon as no coherence remain in the system, the atoms decay and finally reach the ground state.

If the distance $d$ between the nearest neighbors in the lattice is increased, these effects get weaker. More concretely, the superradiant burst decreases and disappears above a certain critical spacing $d_{crit}$, the maximum two-particle coherence is reduced, and the subradiant phase vanishes. For interparticle distances much larger than the atomic transition wavelength, as is the case of the peach (light) curve with $d=2\lambda$, the non-interacting case is recovered. That is, the atomic dipoles do not build up coherences and simply decay exponentially at the fixed rate $\gamma$.

\begin{figure}
\includegraphics[width=\linewidth]{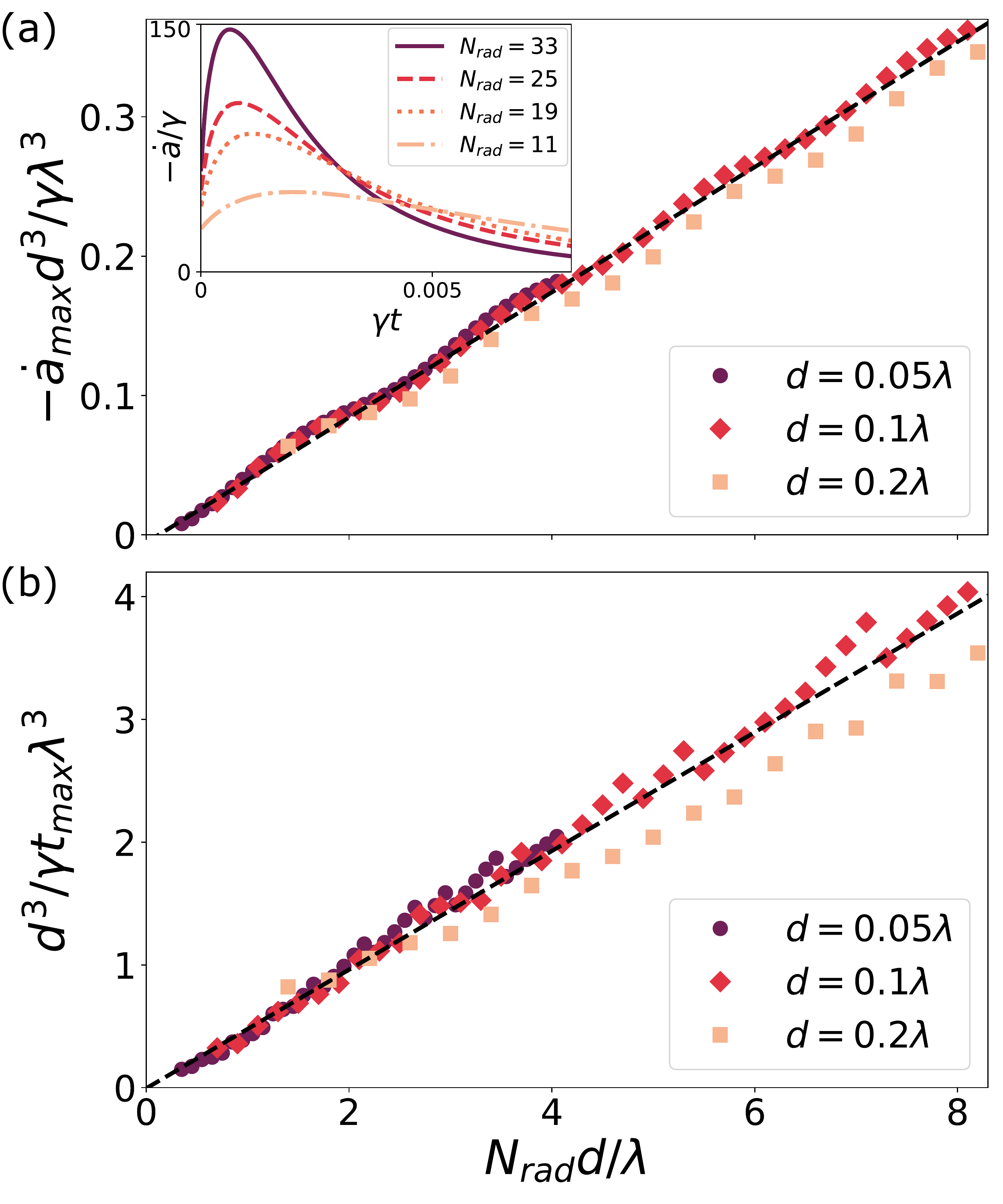}
\caption{\label{fig: 3D_peak} \textbf{Superradiant peak for a three-dimensional array} (a) Maximum emission rate per particle $-\dot{a}_{max}$ multiplied by the dimensionless parameter $d^3/\lambda^3$ versus radius or characteristic length of the sample $N_{rad}d/\lambda$ for different lattice spacings $d$. The inset shows the emitted intensity per particle as a function of time for atomic arrays with fixed $d=0.1\lambda$ and different $N_{rad}$. (b) Time at which the maximum emission occurs versus sample radius and for different spacings. As shown by the black linear fittings, $-\dot{a}_{max}$ scales with $N_{rad}\lambda^2/d^2$ and $t_{max}$ with $d^2/N_{rad}\lambda^2$. We use $\bar{\Gamma}^{(av.)}$ for both panels.}
\end{figure}

The time evolution of the two probe atoms can be further used to characterize the superradiant peak. The inset in Fig.~\ref{fig: 3D_peak}(a) shows the intensity per particle during the burst for three-dimensional arrays of spacing $d=0.1\lambda$ and different numbers of particles along the radial direction $N_{rad}$. The magnitude of the peak $-\dot{a}_{max}$ increases with lattice size and the point at which the emission is maximum $t_{max}$ shifts to earlier times. The exact scaling of both features depends on two quantities: the characteristic length or size of the array, given by $N_{rad} d/\lambda$, and the number of particles within a cubic atomic transition wavelength $\lambda^3/d^{3}$, which corresponds to the density of the sample and coincides with the relevant length scale that appears in the three-dimensional retarded Green's function through the parameter $\xi$ given in Eq.~(\ref{eq: retardedGreen_3D}). For a three-dimensional ensemble, its product results in the optical depth of the medium $\mathcal{O} = N_{rad}\lambda^2/d^2$. As illustrated in Fig.~\ref{fig: 3D_peak}, we find that the maximum emission rate per particle scales linearly with the optical depth $-\dot{a}_{max} \propto \mathcal{O}$, whereas the time at which the maximum emission occurs is inversely proportional to it $t_{max} \propto \mathcal{O}^{-1}$ \cite{intro_gross}. For a spherical sample, the number of particles along its characteristic direction, that is, its radius, scales as $N_{rad} \propto N^{1/3}$, where $N$ is the total amount of atoms in the array. Thus, the total peak intensity emitted by the array scales as $N \times \mathcal{O} \propto N^{4/3}$, well below the typical $N^2$ scaling found in the Dicke limit, where all atoms are contained in a volume much smaller than $\lambda^3$. Note that the optical depth of a sample with a fixed number of atoms depends on the specific shape of the system. Samples with a preferential axis, such as cigar-shaped clouds, have a smaller amount of transverse photon modes \cite{Akkermans} and can therefore attain a quadratic dependence of the pulse intensity with atom number \cite{cigar-shaped}. Additionally, such samples do not scatter photons in all directions, as is the case for spherical arrays or clouds \cite{Isotropic_spherical}, but emit light predominantly along the preferential directions with highest optical depth \cite{Experiment_3,cigar-shaped}.

\begin{figure}
\includegraphics[width=\linewidth]{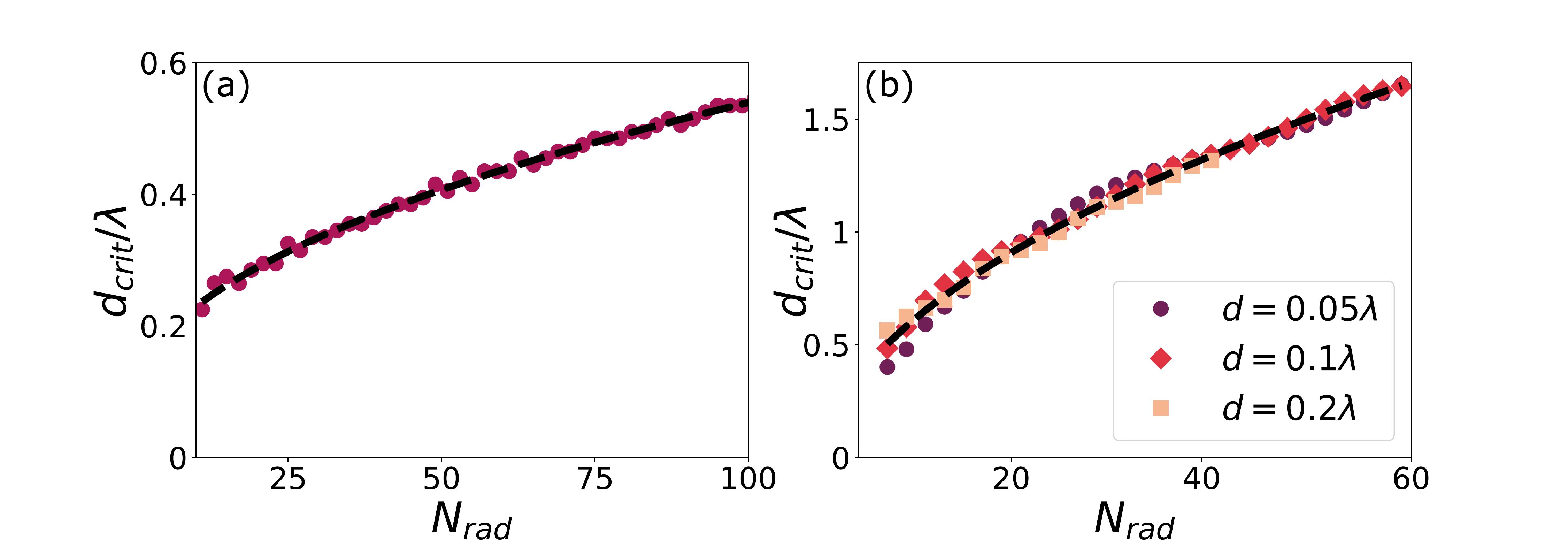}
\caption{\label{fig: phase_diagram_3D} \textbf{Phase diagram for a three-dimensional array} (a) Maximum spacing $d_{crit}$ at which superradiance is sustained as a function of system size $N_{rad}$. A burst is considered to occur if the emitted intensity per particle initially grows, that is, if $-d^2 a/dt^2>0$. (b) Phase diagram estimated from the scaling of the superradiant peak $-\dot{a}_{max}d^2/\lambda^2 \approx f(N_{rad})$. Using the traces in Fig.~\ref{fig: 3D_peak}(a), the critical spacing for a given $N_{rad}$ is obtained as $d_{crit}=d \sqrt{-\dot{a}_{max}}$. The black dashed curves are fittings of the form $d_{crit}=a+b\sqrt{N_{rad}}$.}
\end{figure}

Defining a superradiant burst to occur if the emitted intensity per particle initially increases ($d^2 a/dt^2\!<\!0$), one can obtain the superradiant phase diagram for three-dimensional atomic arrays in Fig.~\ref{fig: phase_diagram_3D}(a). A burst appears below a critical interparticle spacing $d_{crit}$, that is, for a dense enough medium. The spacing $d_{crit}$ depends on the size of the array such that larger samples can sustain superradiance at larger lattice constants. Note that these values are much lower than those reported in other references \cite{superradiantburst_ana,robicheaux_superradiance,Ana_super_new} and have to be understood as a very conservative estimate. This is due to the self-consistent procedure used to compute the cooperative rate $\Gamma$, which considers interactions and cooperativity to be present from the beginning. As a result, the initial decay rate $-\dot{a}(t=0)$ is overestimated and masks the appearance of a burst at finite time if the superradiant peak is not prominent enough. Alternatively, one can obtain a more realistic estimate of the critical spacing by using the scaling of the superradiant peak $-\dot{a}_{max}d^2/\lambda^2 \approx f(N_{rad})$, where $f$ is a liner function of the number of atoms in the radial direction. For a certain $N_{rad}$, $d_{crit}$ corresponds to the spacing that results in $-\dot{a}_{max}^{crit}=1$, that is, $d_{crit}=d \sqrt{-\dot{a}_{max}}$. Figure~\ref{fig: phase_diagram_3D}(b) shows the resulting phase diagram extracted from the traces in Fig.~\ref{fig: 3D_peak}(a), which qualitatively matches those reported in Refs.~\cite{superradiantburst_ana,robicheaux_superradiance,Ana_super_new}. For both phase diagrams, we obtain a critical spacing that scales as $d_{crit}/\lambda \propto \sqrt{N_{rad}}$ \cite{robicheaux_superradiance} (as shown by the black dashed fitted curves), which coincides with recent theoretical predictions \cite{robicheaux_superradiance,Ana_super_new} and benchmarks the validity of our formalism.

Interestingly, the early dynamics (superradiant and subradiant phases) of the atomic array are very similar to those of a three-dimensional homogeneous gas of atoms with the same size and density of particles \cite{AMO_GuindarLin,Hanzhen}, as shown in Fig.~\ref{fig: time_evolution}(a). More specifically, an identical scaling of the superradiant peak and a phase diagram similar to the one shown in Fig.~\ref{fig: phase_diagram_3D} are found for a homogeneous gas. However, the late dynamics differs considerably. While the partially excited, ordered atomic array rapidly decays once the coherences between atoms vanish, the excitation remains in the system much longer in the case of a homogeneous gas of atoms, giving rise to a radiation trapping regime \cite{RadiationTrapping_1,RadiationTrapping_2,RadiationTrapping_3}. When more than half of the excitation has been emitted, that is, $a<\frac{1}{2}$, the three-dimensional Green's function in the medium $\tilde{D}^{ret}_{3D}$ becomes absorbing and its value decays with distance. Thus, the interaction predominantly occurs between nearest neighbors. Even if the average spacing between the atoms in the gas is of the order of a wavelength, there is a non-negligible chance that some atoms are found to be much closer than that. This gives rise to an enhanced interaction in the gas and therefore a larger collective decay rate $\Gamma$, which ultimately suppresses emission according to Eq.~(\ref{eq: equationsmotion}).

\begin{figure}
\includegraphics[width=\linewidth]{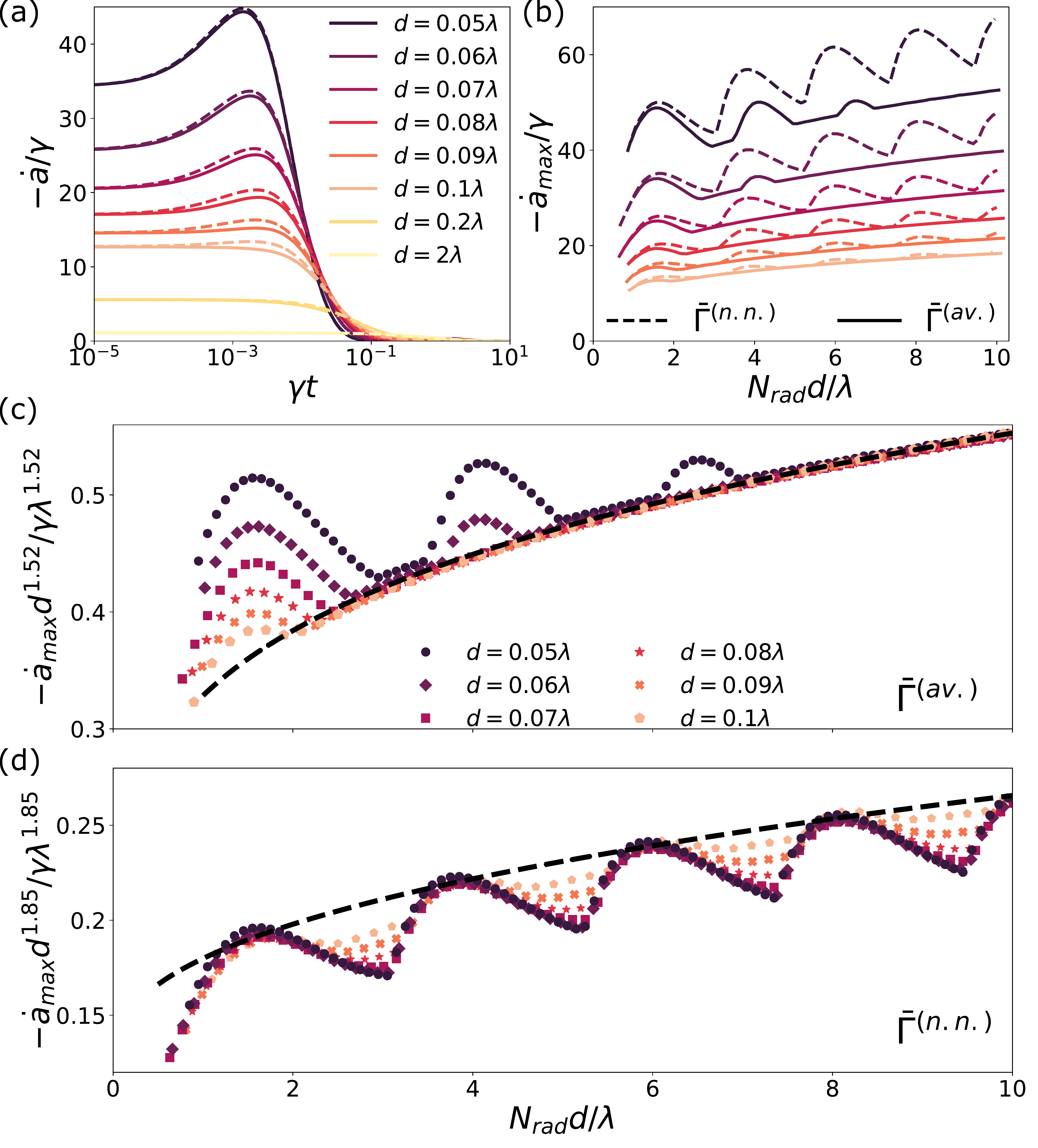}
\caption{\label{fig: 2D_peak} \textbf{Superradiant peak for a two-dimensional array} (a) Intensity per particle $-\dot{a}$ emitted by a two-dimensional atomic array as a function of time for different lattice constants $d$. A circular sample with $N_{rad}=21$ particles in the radial direction ($317$ atoms in total) is considered. (b) Maximum intensity per particle $-\dot{a}_{max}$ for samples with different $N_{rad}$ and $d$. The same color scale is used in all panels. In (a) and (b) the dashed lines correspond to the two-atom cooperative decay rate $\bar{\Gamma}^{(n.n.)}$, whereas the solid lines represent the result for $\bar{\Gamma}^{(av.)}$ (see Appendix~\ref{appendix: decayrates}). (c) Scaling of the superradiant peak resulting from $\bar{\Gamma}^{(av.)}$. The black dashed curve corresponds to a power-law fit of the form $-\dot{a}_{max} (d/\lambda)^{1.52} \propto (N_{rad}d/\lambda)^{0.23}$. An equally good fitting can be obtained with the logarithmic function $-\dot{a}_{max} (d/\lambda)^{1.52} \propto \log(N_{rad}d/\lambda)$. (d) Scaling of the superradiant peak resulting from $\bar{\Gamma}^{(n.n.)}$. The black dashed curve corresponds to a fit of the form $-\dot{a}_{max} (d/\lambda)^{1.85} \propto (N_{rad}d/\lambda)^{0.4}$. }
\end{figure}

\section{Two-dimensional atomic arrays}

For two-dimensional atomic arrays, that is, ensembles of atoms lying on a plane, the time evolution of the average upper-level population $a$ and the two-level coherence $\rho_{eg,ge}$ presents the same three regimes as the three-dimensional case in Fig.~\ref{fig: time_evolution}(a). However, the two-dimensional superradiant burst is weaker, the subradiant phase is less prominent, and both collective effects emerge only at lower interparticle spacings. This is consistent with the fact that three-dimensional lattices are better packed geometries that contain many more particles within a cubic transition wavelength and therefore exhibit stronger cooperative effects. Unlike in the three-dimensional case, the properties of the superradiant burst of two-dimensional arrays depend on the specific way the two-atom cooperative decay rate is computed or, equivalently, on the position of the two probe atoms. Figure~\ref{fig: 2D_peak}(a) depicts the emission rate per atom $-\dot{a}$ for small samples ($N_{rad}=21$) of various spacings and demonstrates the appearance of a superradiant burst at low enough $d$, while Fig.~\ref{fig: 2D_peak}(b) shows the maximum emission rate $-\dot{a}_{max}$ as a function of system size $N_{rad}d/\lambda$. In both cases, the solid lines represent the results obtained with the averaged collective decay rate $\bar{\Gamma}^{(av.)}$, whereas the dashed lines correspond to the dynamics in the case where the probe atoms are nearest neighbors, computed with $\bar{\Gamma}^{(n.n.)}$. We find that the peak intensities obtained with $\bar{\Gamma}^{(n.n.)}$ are generally larger than those corresponding to $\bar{\Gamma}^{(av.)}$. This occurs because the two-atom collective decay rate $\bar{\Gamma}(\vec{r}_1,\vec{r}_2)$ decreases with the distance between probe atoms $|\vec{r}_2-\vec{r}_1|$ (see Appendix~\ref{appendix: 2D array}), which ultimately reduces the coherence $\rho_{eg,ge}$ built in the system and consequently the strength of the cooperative effects. 
As shown in Appendix~\ref{appendix: 2D array}, gradually increasing $|\vec{r}_2-\vec{r}_1|$ when computing $\bar{\Gamma}(\vec{r}_1,\vec{r}_2)$ results in a transition from $\bar{\Gamma}^{(n.n.)}$ to $\bar{\Gamma}^{(av.)}$.

Additionally, Fig.~\ref{fig: 2D_peak}(b) shows that $-\dot{a}_{max}$ does not increase monotonically with sample size, but oscillates with period $2\lambda$. That is, a maximum (or minimum) is reached every time the radius of the atomic array increases by one atomic transition wavelength. This behavior arises from the oscillating nature of the retarded Green's function in the medium, which results in constructive and destructive interference between the different ``shells" of the array when computing the cooperative decay rates $\Gamma$ and $\bar{\Gamma}$. Note that these oscillations also appear in three-dimensional arrays [see Fig.~\ref{fig: 3D_peak}(a)], although the effect is much weaker due to the strongly amplifying nature of the three-dimensional medium.

The functional form of $-\dot{a}_{max}$ can be obtained by appropriately scaling the emission axis. Figure~\ref{fig: 2D_peak}(c) shows that the maximum emission rate per particle obtained with $\bar{\Gamma}^{(av.)}$ scales as $-\dot{a}_{max} \propto N_{rad}^{0.23} (\lambda/d)^{1.29}$. As shown in Appendix~\ref{appendix: 2D array}, a similar scaling is obtained from the minima of $-\dot{a}_{max}$ computed with $\bar{\Gamma}^{(n.n)}$. As for the maxima, the traces in Fig.~\ref{fig: 2D_peak}(d) result in a power-law scaling of the form $-\dot{a}_{max} \propto N_{rad}^{0.4} (\lambda/d)^{1.45}$. Noting that the peak intensity for pairs of probe atoms separated by more than one lattice site ranges between the values obtained with $\bar{\Gamma}^{(n.n.)}$ and $\bar{\Gamma}^{(av.)}$ (see Appendix~\ref{appendix: 2D array}) and using the fact that $N_{rad} \propto N^{1/2}$ in two-dimensional lattices, we can conclude that the total peak intensity radiated by the array scales as a power law $N^{\alpha}$ with exponent $\alpha \in \{1.115,1.2\}$.
As expected, we obtain a smaller exponent than that of three-dimensional arrays (where cooperative effects are stronger) and nonextended systems (where the Dicke limit holds).

 We finally note that the initial slope of the total radiation [$-N \dot{a}(t=0)$] was recently found to scale in two-dimensional arrays with the logarithm of $N_{rad}$ \cite{robicheaux_superradiance}. Motivated by this result, we find that the minima of $-\dot{a}_{max}$ computed with $\bar{\Gamma}^{(n.n.)}$, as well as the traces in Fig.~\ref{fig: 2D_peak}(c) obtained with $\bar{\Gamma}^{(av.)}$, are also compatible with a logarithmic scaling. That is, both the logarithmic function and the power law overlap for systems of length up to ten times the natural transition wavelength. This corresponds to arrays of $100 \times 100$ atoms in the case of $d=0.1\lambda$, well beyond the size that has been experimentally realized in lattices of cold atoms with subwavelength spacing \cite{OpticalLAttice_1}. 

\section{Conclusion and outlook}

We have analyzed the many-body dynamics of closely spaced and dipole-coupled atomic arrays by means of a reduced two-atom master equation that captures correlations with the rest of the ensemble. As opposed to the formalism used in Refs.~\cite{superradiantburst_ana,Ana_super_new,robicheaux_superradiance}, which perfectly captures the photon emission at zero time, our method overestimates the initial cooperative effects in the array and consequently does not provide accurate estimates of the superradiant phase diagrams. However, it satisfactorily captures the mid- and long-term behaviors of the atomic system. This allowed us not only to demonstrate the appearance of superradiance and subradiance below a critical spacing, but also to characterize the scaling of the superradiant peak for three-dimensional and two-dimensional atomic arrays. In particular, we showed that the total intensity in extended samples scales with a lower exponent than in the ideal Dicke case, where all atoms are contained within a cubic transition wavelength, and found that three-dimensional arrays present a larger exponent than their two-dimensional counterparts, consistent with the notion that three-dimensional lattices are better packed geometries that exhibit stronger cooperative effects. We additionally showed that the figures of merit for the superradiant burst of ordered arrays and homogeneous gases of atoms are similar and identified significant differences in the late-time dynamics of both systems. As opposed to arrays, homogeneous gases can sustain radiation trapping once the atomic coherences vanish due to a nonzero probability of finding two atoms at distances much lower than the average interparticle spacing.

The collective phenomena studied in this paper may be experimentally realized in a wide variety of platforms, ranging from ultracold atoms trapped in optical lattices \cite{OpticalLAttice_1,3Dlattice_JunYe} and tweezers \cite{Tweezer_2,Tweezer_3} to condensed-matter systems such as quantum emitters in two-dimensional materials \cite{2D_mat_1,2dmat_2} or color centers in bulk crystals \cite{vacancy_1,vacancy_2,vacancy_3}. Further, this work could be extended by adding a classical driving field, which may elucidate the behavior of arrays in other regimes of the multiexcitation sector for which theoretical and numerical understanding is still very limited \cite{robicheaux_superradiance,multi_exc_molmer,2_exc_Ana}. Also, the effective two-atom description of the many-body problem can be potentially leveraged to study other systems or reservoirs by appropriately modifying the Green's function of the medium \cite{OtherReservoirs}.

We thank Hanzhen Ma for insightful conversations about the effective two-atom model used to describe the atomic arrays. We also acknowledge valuable discussions with Ana Asenjo-Garcia and Stuart J. Masson. This work was supported by the NSF through the CUA Physics Frontier Center and through PHY-1912607, as well as by the AFOSR through Grant No. FA9550-19-1-0233.
O.R.-B. acknowledges support from Fundació Bancaria “la Caixa” (Grant No. LCF/BQ/AA18/11680093).

\appendix

\section{Retarded Green's function in the medium}
\label{appendix: greensFunction}

\begin{figure}[b]
\includegraphics[width=\linewidth]{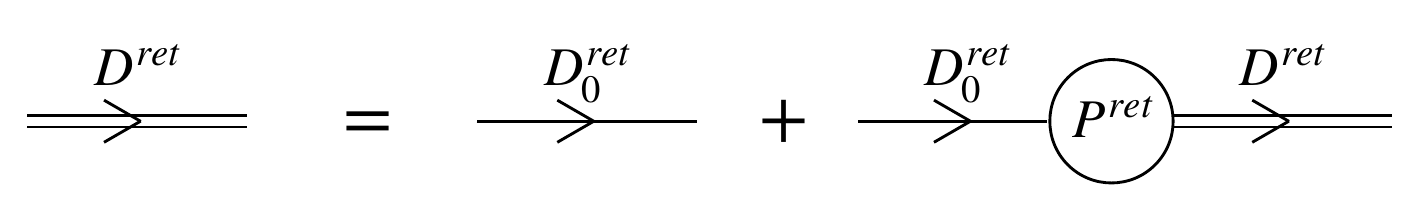}
\caption{\label{fig: dyson_graphical} Graphical representation of the Dyson equation. The retarded Green's function inside the medium $D^{ret}$ is generated by the free-space Green's function $D^{ret}_0$ and the polarization source function $P^{ret}$.}
\end{figure}

The retarded Green's function in the medium is obtained from the free-space Green's function $D^{ret}_0$ and the polarization source function $P^{ret}$ using the Dyson equation formalism \cite{Dyson,Fleishbauer}, which is graphically represented in Fig.~\ref{fig: dyson_graphical}. It can be formally written as

\begin{align}
\label{eq: dyson_general}
    D^{ret}_{\alpha \beta}(\vec{r}_1,t_1;\vec{r}_2,t_2) =D^{ret}_{0\alpha \beta}(\vec{r}_1,t_1;\vec{r}_2,t_2) - \int_{-\infty}^\infty dt_1' \int_{-\infty}^\infty dt_2' \nonumber \\
    \times \int_V d^3 \vec{r'_1} D^{ret}_{0\alpha \mu}(\vec{r}_1,t_1;\vec{r'_1},t_1') P^{ret}_{\mu \nu} (\vec{r'_1} ;t_1',t_2') D^{ret}_{\nu \beta}(\vec{r_1'},t_2';\vec{r}_2,t_2),
\end{align}
where $\alpha$ and $\beta$ represent the components of the Green's tensor.

The polarization function $P^{ret}$ is given by the correlation function of dipole operators of non-interacting atoms, which can be computed using the quantum regression theorem \cite{AMO_GuindarLin}. Using a continuum approximation, $P^{ret}$ can be expressed as
\begin{equation}
\label{eq: source_function}
    P^{ret}(\vec{r},t)= \frac{\wp ^2}{\hbar ^2} \frac{1}{d^\mathcal{D}} \frac{2a(\vec{r},t)-1}{\gamma/2+\Gamma},
\end{equation}
where $d$ is the lattice constant, $a$ is the average upper-level population of the two probe atoms, $\Gamma$ is the cooperative decay rate, and $\mathcal{D}$ represents the dimensionality of the array, that is, $\mathcal{D}=3$ for three-dimensional lattices and $\mathcal{D}=2$ for two-dimensional ones.

Equation~(\ref{eq: dyson_general}) can be solved in Fourier space if a series of approximations are done \cite{Fleishbauer}. First, we extend the spatial integral to infinity. Second, the spatial dependence of the source function is replaced by $\vec{r}_2$. Finally, we make use of the Markov approximation to only keep the slow time dependence of the source function and assume that it depends on the time difference $t_1'-t_2'$. We can then Fourier transform with respect to space $\vec{x}=\vec{r}_1-\vec{r}_2$ and time $\tau=t_1-t_2$ to obtain
\begin{equation}
\label{eq: greenmedium_tensor}
    \mathbf{\tilde{\tilde{D}}^{ret}}(\vec{q},\omega;t)= \left[\mathbf{1}+\mathbf{\tilde{\tilde{D}}_0^{ret}}(\vec{q},\omega) \mathbf{\tilde{P}^{ret}}(\omega;t) \right]^{-1} \mathbf{\tilde{\tilde{D}}_0^{ret}}(\vec{q},\omega) ,
\end{equation}
where $\mathbf{\tilde{\tilde{D}}^{ret}}$ and $\mathbf{\tilde{P}^{ret}}$ are $3 \times 3$ matrices and $\mathbf{1}$ is the identity matrix.

The free-space retarded Green's function in real space is 
\begin{equation}
        \tilde{D}_{0\alpha\beta}^{ret}(\vec{x},k_0)= -\frac{i\hbar}{4\pi \epsilon_0} \left(k_0^2\delta_{\alpha\beta}+ \frac{\partial^2}{\partial x_{\alpha} \partial x_{\beta}} \right) \frac{e^{-i k_0 r}}{r} ,
\end{equation}
where $r=|\vec{x}|$.  If the medium is randomly polarized, one can apply the polarization average $\langle \wp_\alpha \wp_\beta \rangle = \frac{1}{3} \delta_{\alpha \beta}$. This is equivalent to performing the orientation average $x_\alpha x_\beta/r^2 \rightarrow \langle x_\alpha x_\beta / r^2 \rangle = \frac{1}{3} \delta_{\alpha \beta}$. The Green's function then becomes a spherical tensor with components
\begin{equation}
\label{eq: freespace_real_approx}
    \tilde{D}_0^{ret}(\vec{x},k_0) 
    = -\frac{i\hbar k_0^2}{6\pi \epsilon_0}  \frac{e^{-i k_0 r}}{r},
\end{equation}
with $k_0=2\pi/\lambda$. The spherical nature of the problem now simplifies Eq.~(\ref{eq: greenmedium_tensor}) to the scallar equation
\begin{eqnarray}
\label{eq: retarded_scalar}
    \tilde{\tilde{D}}^{ret}(\vec{q},k_0)= \frac{1}{\left[\tilde{\tilde{D}}^{ret}_{0}(q,k)\right]^{-1}+P^{ret}}. 
\end{eqnarray}
The $\tilde{\tilde{D}}^{ret}_{0}(\vec{q},k_0)$ is obtained by Fourier transforming Eq.~(\ref{eq: freespace_real_approx}) and it therefore depends on the dimensionality of the sample.  

\subsection{Three-dimensional sample}

For a three-dimensional sample, the free-space Green's function in momentum space is 
\begin{equation}
    \tilde{\tilde{D}}^{ret}_{0}(\vec{q},k_0)= -\frac{2i\hbar k_0^2}{3\epsilon_0} \frac{1}{q^2-k_0^2+2i k_0 \epsilon},
\end{equation}
where the small positive constant $\epsilon$ moves the pole at $q=k_0$ to the lower half of the complex plane. Plugging this result in Eq.~(\ref{eq: retarded_scalar}), performing the inverse Fourier transform with respect to $\vec{q}$, and inserting the explicit form of the source function given by Eq.~(\ref{eq: source_function}), we finally obtain Eq.~(\ref{eq: retardedGreen_3D}) of the main text \cite{Fleishbauer}
\begin{align}
    \tilde{D}_{3D}^{\textit{ret}}(r) & =-\frac{i\hbar k_0^2}{6 \pi \epsilon_0}\frac{e^{-i k_0 r} e^{\xi r}}{r}, \nonumber \\
    \xi &= \gamma \frac{2a-1}{\gamma/2+\Gamma} \frac{\pi}{k_0^2 d^3}, 
\end{align}
where we have defined the spontaneous decay rate $\gamma=\wp^2 k_0^3/3\pi \epsilon_0 \hbar$.

Thus, the Green's function in a three-dimensional medium oscillates with period $\lambda$. For a predominantly excited medium such that the average upper-level population $a>0.5$, $\xi$ is positive and $\tilde{D}_{3D}^{\textit{ret}}$ exponentially increases with distance. In this regime, the medium is amplifying. For $a<0.5$, the medium becomes absorbing and $\tilde{D}_{3D}^{\textit{ret}}$ decreases with distance.

\subsection{Two-dimensional sample}

We assume that the atomic sample is located in the $xy$ plane such that $z=0$ for all atoms. Then, the Fourier transform is carried out only over $x$ and $y$ and the retarded free-space Green's function in momentum space is
\begin{equation}
        \tilde{\tilde{D}}^{ret}_{0}(\vec{q},z=0,k_0)=-\frac{i \hbar k_0}{3 \epsilon_0} \frac{1}{\sqrt{ q^2/k_0^2 -1+2i\epsilon/k_0}},
\end{equation}
where the momentum is now defined in two dimensions $\vec{q}=(k_x,k_y)$ and $q=\sqrt{q_x^2+q_y^2}$. Again, a small positive constant $\epsilon$ is introduced.

\begin{figure}[t]
\includegraphics[width=\linewidth]{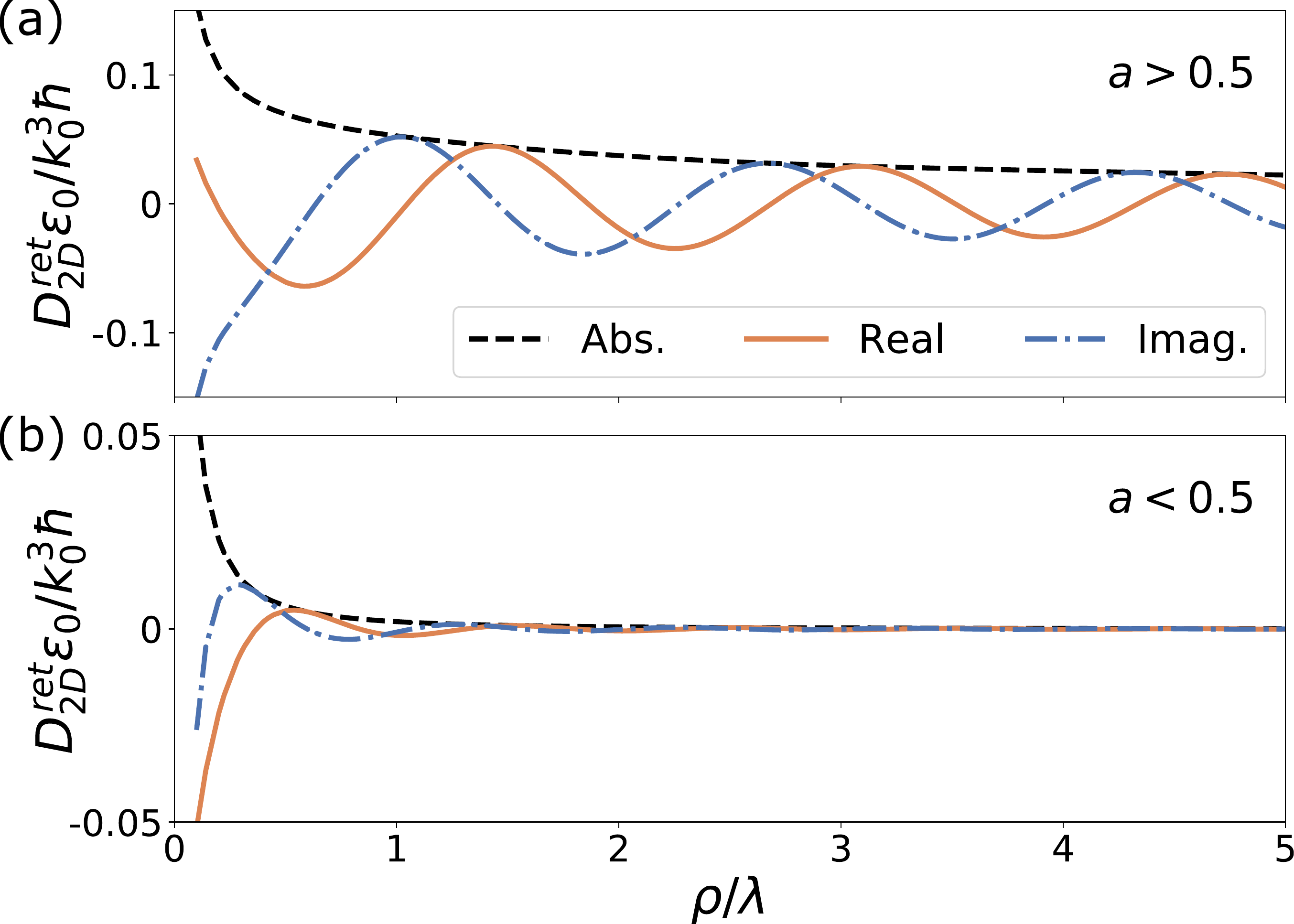}
\caption{\label{fig: Dret_2D} \textbf{Two-dimensional retarded Green's function} The retarded Green's function in a two-dimensional medium $\tilde{D}^{ret}_{2D} \epsilon_0 / k_0^3 \hbar$ is plotted as a function of distance for (a) $a>0.5$ and $\chi=0.8$ and (b) $a<0.5$ and $\chi=-0.8$. The orange solid and blue dash-dotted traces correspond to the real and imaginary parts respectively, whereas the black dashed line represents the absolute value. A spacing of $d=0.1\lambda$ is considered.}
\end{figure}

From Eqs.~(\ref{eq: retarded_scalar}) and~(\ref{eq: source_function}), it follows that
\begin{equation}
        \tilde{\tilde{D}}^{ret}_{2D}(\vec{q},k_0)= -\frac{i \hbar k_0}{3 \epsilon_0} \frac{1}{\sqrt{ q^2/k_0^2-1+2i\epsilon/k_0}-i \chi},
\end{equation}
where we have defined the parameter 
\begin{equation}
    \chi=\frac{\gamma}{\gamma/2+\Gamma} \frac{\pi}{k_0^2 d^2} (2a-1),
\end{equation}
which depends on the state of the two probe atoms and the lattice constant of the array. Performing the inverse Fourier transform, we obtain
\begin{align}
\label{eq: retarded_2d_appendix}
        \tilde{D}^{ret}_{2D}(\rho,z=0) = \frac{1}{ \left( 2\pi \right) ^2} \int_{-\infty}^\infty dq_x \int_{-\infty}^{\infty} dq_y \tilde{\tilde{D}}^{ret}(\vec{q},k_0) e^{-i \vec{q}\vec{r}}
     \nonumber \\
     = \frac{1}{ \left( 2\pi \right) ^2}
    \int_0^{2\pi} d\theta \int_{0}^{\infty} q dq  \tilde{\tilde{D}}^{ret}(\vec{q},k_0) e^{-i q \rho \cos \theta} \nonumber \\
       =-\frac{i \hbar k_0}{6 \pi \epsilon_0}  \int_{0}^{\infty} dq \frac{q J_0(q\rho)}{\sqrt{ q^2/k_0^2-1+2i\epsilon/k_0}-i \chi},
\end{align}
where $J_0$ is the zeroth-order Bessel function of the first kind and $\rho=\sqrt{x^2+y^2}$ is the distance between two points on the $xy$ plane.

The integral is performed numerically for the discrete set of distances $\rho$ that appear in an atomic array. That is, given a lattice with spacing $d$, one needs to consider $\rho=d \sqrt{n_x^2+n_y^2}$, where $n_x$ and $n_y$ are integers. Note also that Eq.~(\ref{eq: retarded_2d_appendix}) has a pole at $q=\pm k_0 \sqrt{1- \chi^2}$. The small constant $\epsilon$ therefore ensures the convergence of the integral when the pole is located in the real axis.

As shown in Fig.~\ref{fig: Dret_2D}, $\tilde{D}^{ret}_{2D}(\rho)$ oscillates and its absolute value decays with distance for all values of $\chi$. That is, the medium is absorbing for all average upper-level populations $a$.

\section{Cooperative decay rates}
\label{appendix: decayrates}

After tracing out the degrees of freedom of the electromagnetic field and the $N-2$ nonselected atoms, the resulting master equation for the reduced system, and therefore the equations of motion given by Eq.~(\ref{eq: equationsmotion}), depends on the cooperative decay rates $\Gamma$ and $\bar{\Gamma}$. These quantities can be expressed in terms of the two-time cumulants of the field operators and are therefore related to the Green's function in the atomic medium. As shown in Refs.~\cite{AMO_GuindarLin,Hanzhen}, one can find the closed-form expressions 
\begin{widetext}
\begin{align}
\label{eq: collectivedecayrates_1}
\Gamma(\vec{r}) &= \frac{\wp^4}{\hbar^4} \sum_{\vec{x}} \frac{2a}{\gamma/2+\Gamma} \left| \tilde{D}^{ret} (\vec{r}-\vec{x}) \right|^2 + \frac{\wp^4}{\hbar^4} \sum_{\vec{x}_1} \sum_{\vec{x}_2} \frac{2\rho_{eg,ge}}{\gamma/2+\Gamma} \tilde{D}^{ret} (\vec{r}-\vec{x}_1) \tilde{D}^{*ret} (\vec{r}-\vec{x}_2), \nonumber \\
\bar{\Gamma}(\vec{r}_1,\vec{r}_2) &=  \frac{\wp^4}{\hbar^4} \sum_{\vec{x}} \frac{2a}{\gamma/2+\Gamma} \tilde{D}^{ret} (\vec{r}_1-\vec{x}) \tilde{D}^{*ret} (\vec{r}_2-\vec{x}) + \frac{\wp^4}{\hbar^4} \sum_{\vec{x}_1} \sum_{\vec{x}_2} \frac{2\rho_{eg,ge}}{\gamma/2+\Gamma} \tilde{D}^{ret} (\vec{r}_1-\vec{x}_1) \tilde{D}^{*ret} (\vec{r}_2-\vec{x}_2),
\end{align}
\end{widetext}
where the summation is carried out over all the atoms of the system, located at positions $\vec{x}$. The specific form of the retarded Green's function depends on the dimensionality of the lattice and the collective decay rates depend on the positions of the two probe atoms $\vec{r}_1$ and $\vec{r}_2$. Here $\Gamma$ and $\bar{\Gamma}$ can be understood as the one-atom and two-atom cooperative decay rates, respectively. That is, $\Gamma$ appears in the reduced master equation through terms involving raising and lowering operators of one probe atom only (e.g., $\sigma_1 \sigma_1^\dagger$), while $\bar{\Gamma}$ corresponds to terms that involve both probe atoms (e.g., $\sigma_1 \sigma_2^\dagger$). Note also that the decay rates for a homogeneous gas can be obtained by the replacement $\sum_{\vec{x}} \rightarrow \mathcal{N} \int_V d^3\vec{x}$, where $\mathcal{N}$ denotes the density of the medium \cite{AMO_GuindarLin,Hanzhen}. For clarity, we define $\Gamma=\Gamma_1+\Gamma_2$ and $\bar{\Gamma}=\bar{\Gamma}_1+\bar{\Gamma}_2$, where the subindices indicate the first and second terms of both collective decay rates.

We here assume that the spatial dependence of the atomic variables is much weaker than that of the field correlations, which rapidly oscillate according to $\tilde{D}^{ret}$. We thus describe the atomic system with the averaged variables $a$, $n$, and $\rho_{eg,ge}$. Physically, this assumption amounts to neglecting retardation effects of the electromagnetic field (as well as the edge effects that might arise from the boundaries of finite-size systems). Additionally, we consider that the one-atom cooperative decay rate can be approximated as $\Gamma \approx \Gamma(\vec{r}=\vec{0})$, consistent with the fact that the majority of the atoms are deep inside the array for large enough samples. Similarly, $\bar{\Gamma}_2$ is approximated as $\sum_{\vec{x}_1} \tilde{D}^{ret} (\vec{r}_1-\vec{x}_1) \sum_{\vec{x}_2} \tilde{D}^{*ret} (\vec{r}_2-\vec{x}_2) \approx \sum_{\vec{x}} \left| \tilde{D}^{ret} (\vec{x}) \right|^2$ and is therefore equal to $\Gamma_2$. However, $\bar{\Gamma}_1$ strongly depends on the choice of $\vec{r}_1$ and $\vec{r}_2$, as the addends in $\sum_{\vec{x}} \tilde{D}^{ret} (\vec{r}_1-\vec{x}) \tilde{D}^{*ret} (\vec{r}_2-\vec{x})$ interfere differently depending on the exact value of both quantities. In order to account for this dependence and obtain the behavior representative of the whole ensemble, we consider and compare different ways of computing $\bar{\Gamma}_1$.

\textit{Case (i):} $\bar{\Gamma}_1^{(pair)} = \bar{\Gamma}_1(\vec{r}_1=\vec{0},\vec{r}_2)$. This is the decay rate for a specific pair of atoms located at positions $\vec{r}_1=0$ and $\vec{r}_2$. We label the specific case of nearest neighbors, where $|\vec{r}_2-\vec{r}_1|=d$, as $\bar{\Gamma}_1^{(n.n)}$.

\textit{Case (ii):} $\bar{\Gamma}_1^{(mean)} = \frac{1}{N} \sum_{\vec{r}_2 \neq \vec{0}} \bar{\Gamma}_1(\vec{r}_1=\vec{0},\vec{r}_2)$. This is the average or arithmetic mean over all possible atom pairs, considering that one of the atoms is at the center of the array.

\textit{Case (iii):} $\bar{\Gamma}_1^{(av.)} = \frac{1}{N} \frac{\wp^4}{\hbar^4} \frac{2a}{\gamma/2+\Gamma} \left| \sum_{\vec{x}} \tilde{D}^{ret} (\vec{x}) \right|^2$. This is the alternative average introduced in Ref.~\cite{AMO_GuindarLin}, which results from adding an extra summation $\frac{1}{N} \sum_{\vec{x}_2}$ and thus separating the $\vec{x}$ dependence into two different variables $\vec{x}_1$ and $\vec{x}_2$. While $\bar{\Gamma}_1^{(av.)}$ provides similar values to $\bar{\Gamma}_1^{(mean)}$ (see Appendix~\ref{appendix: 3D array} and Appendix~\ref{appendix: 2D array}), it is faster to compute and therefore allows us to study larger lattices.

The results presented in the main text are obtained using $\bar{\Gamma}_1^{(av.)}$ in the case of the three-dimensional arrays and $\bar{\Gamma}_1^{(av.)}$ and $\bar{\Gamma}_1^{(n.n.)}$ in the case of two-dimensional lattices. Also, the differences between the various ways of computing $\bar{\Gamma}_1$ for both dimensionalities are discussed in Appendix~\ref{appendix: 3D array} and Appendix~\ref{appendix: 2D array}, respectively.

\section{Three-dimensional array}
\label{appendix: 3D array}

\begin{figure}[t]
\includegraphics[width=\linewidth]{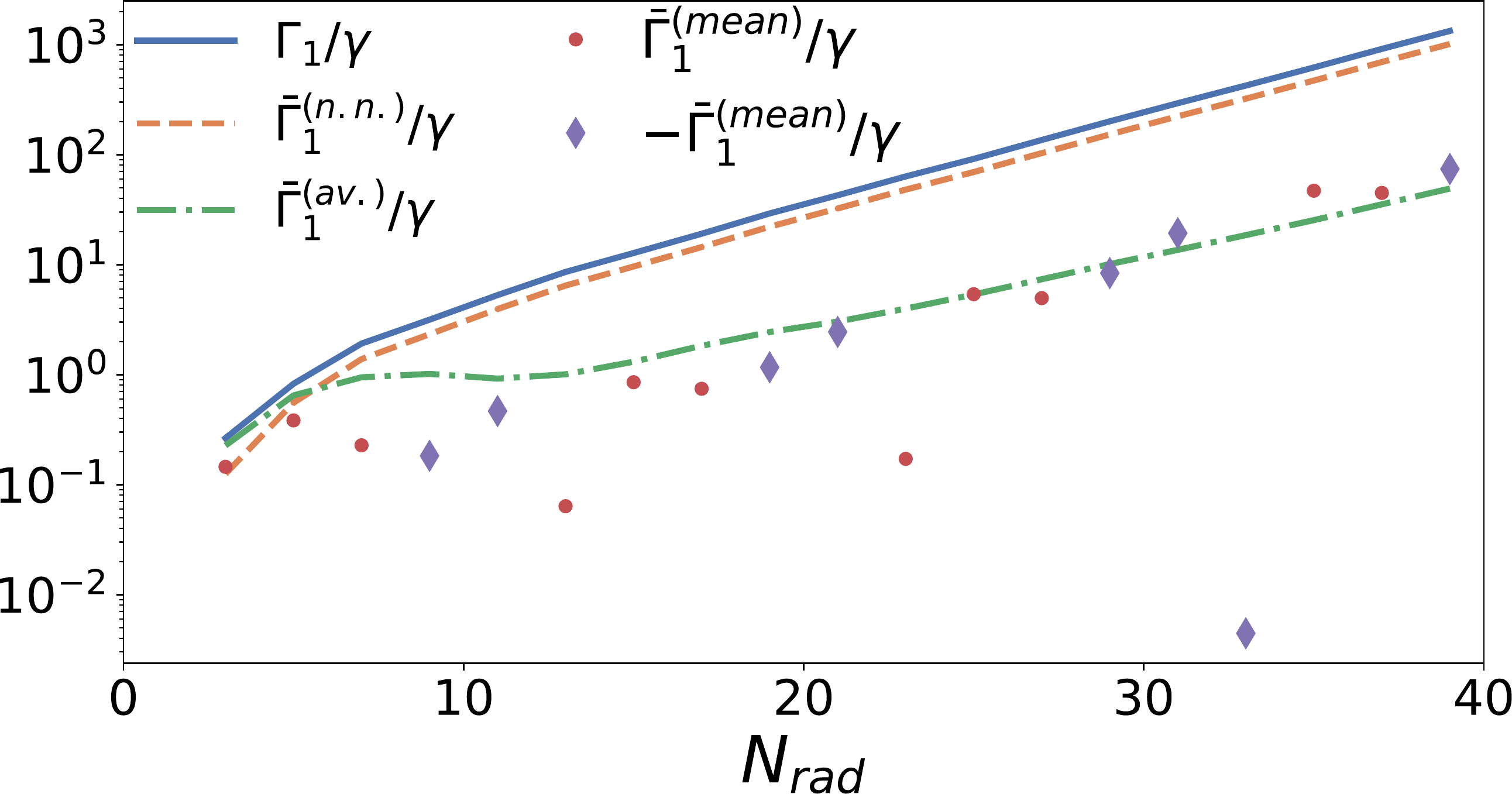}
\caption{\label{fig: 3D_Gbar_different} \textbf{$\bar{\Gamma}_1$ in a three-dimensional array} $\Gamma_1/\gamma a$ (blue solid line) and $\bar{\Gamma}_1/\gamma a$ as a function of the number of atoms in the radial direction $N_{rad}$ for an array with $d=0.2\lambda$ and $\Gamma=10\gamma$. The different colors correspond to the different ways of computing $\bar{\Gamma}_1$: $\bar{\Gamma}_1^{(n.n.)}/\gamma a$ as an orange dashed line, $\bar{\Gamma}_1^{(av.)}/\gamma a$ as a green dash-dotted line, and $\bar{\Gamma}_1^{(mean)}/\gamma a$ and $-\bar{\Gamma}_1^{(mean)}/\gamma a$ as red circles and purple diamonds, respectively.}
\end{figure}

Figure~\ref{fig: 3D_Gbar_different} shows the different values of $\bar{\Gamma}_1$ obtained for a three-dimensional square lattice with spherical shape. For an individual pair of atoms at positions $\vec{r}_1$ and $\vec{r}_2$, it is simply proportional to $\sum_{\vec{x}} \tilde{D}^{ret} (\vec{r}_1-\vec{x}) \tilde{D}^{*ret} (\vec{r}_2-\vec{x})$. Given that the retarded Green's function oscillates with distance, both factors overlap in different ways depending on the relative position of $\vec{r}_1$ and $\vec{r}_2$. If both atoms are at positions such that $\tilde{D}^{ret} (\vec{r}_1)$ and $\tilde{D}^{*ret} (\vec{r}_2)$ have the same sign, the retarded Green's functions overlap in phase and the addends add up constructively. Also, the farther away both atoms are, the smaller the resulting sum is. However, if the signs of $\tilde{D}^{ret} (\vec{r}_1)$ and $\tilde{D}^{*ret} (\vec{r}_2)$ differ, the Green's function at the atomic positions have opposite phases and the resulting $\bar{\Gamma}$ can become negative. This represents a nonphysical scenario in which $\Gamma$ can in turn also become negative during the time evolution, probably due to an overestimation of the phase coherence over distance. 

For $\bar{\Gamma}_1^{(n.n.)}$, there is an almost perfect constructive overlap and the resulting decay rate (orange dashed trace) is always positive and close to $\Gamma_1 = \bar{\Gamma}_1^{(pair)}(\vec{r}_1=\vec{r}_2)$ (blue solid trace). Both $\bar{\Gamma}_1^{(mean)}$ and $\bar{\Gamma}_1^{(av.)}$ represent averages over different atom pairs and their absolute values are therefore smaller than those of $\Gamma_1$ or $\bar{\Gamma}_1^{(n.n.)}$. The arithmetic mean $\bar{\Gamma}_1^{(mean)}$ additionally results in regions with positive (red circles) and negative (purple diamonds) decay rates, which alternate every time the sample size increases by $\lambda$. Interestingly, $\bar{\Gamma}_1^{(av.)}$ is always positive and is close to the absolute value of $\bar{\Gamma}_1^{(mean)}$. Note also that $\Gamma_1$, $\bar{\Gamma}_1^{(n.n.)}$ and $\bar{\Gamma}_1^{(av.)}$ only involve a summation over the $N \approx N_{rad}^3$ lattice sites of the array, whereas $\bar{\Gamma}_1^{(mean)}$ contains two nested summations, which increases the number of operations quadratically and largely reduces the maximum array size that can be numerically simulated.

\begin{figure}
\includegraphics[width=\linewidth]{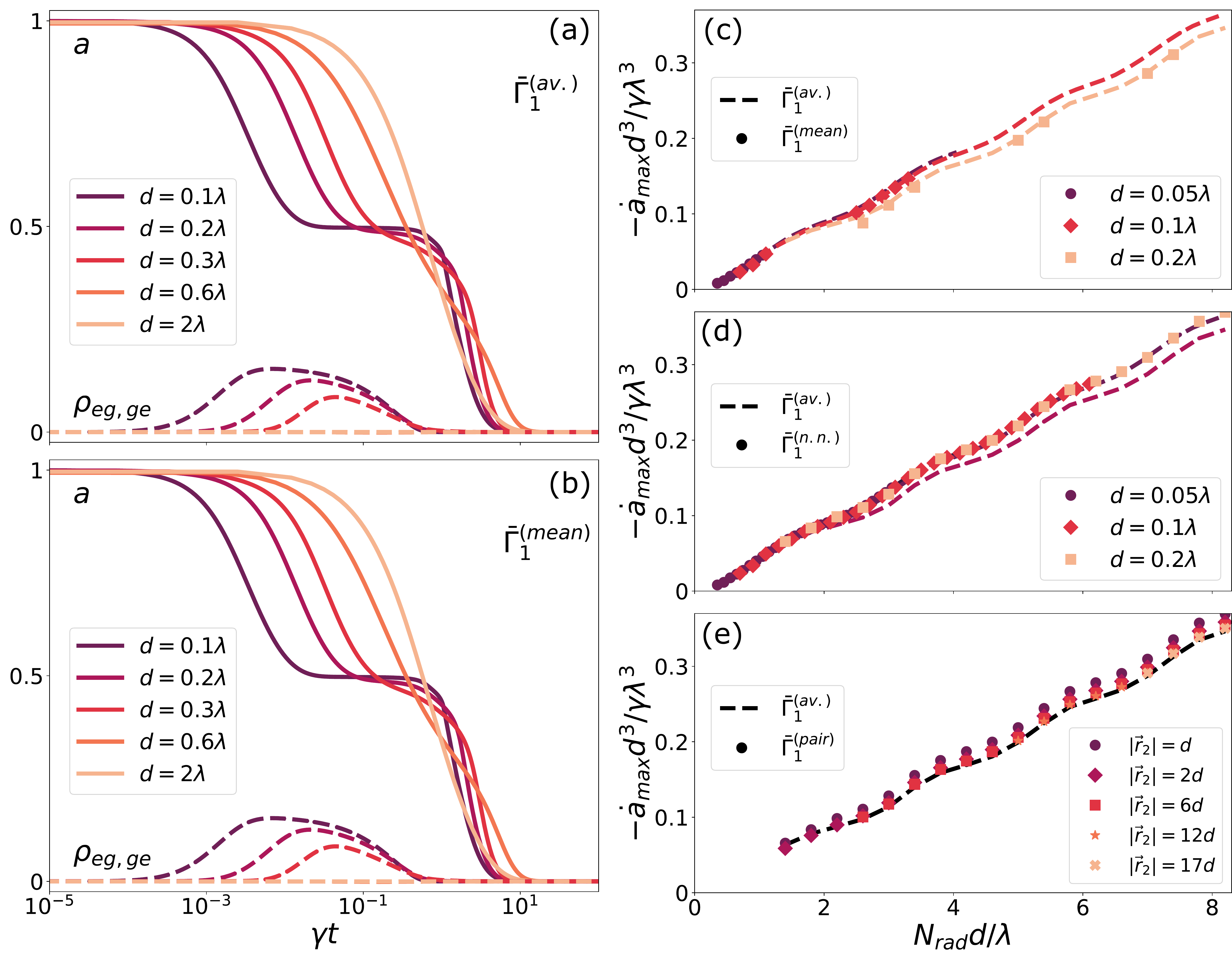}
\caption{\label{fig: 3D_differentmethods} \textbf{Result comparison for three-dimensional arrays} (a)-(b) Average upper-level population $a$ (solid lines) and two-atom coherence $\rho_{eg,ge}$ (dashed lines) as a function of time for a spherical three-dimensional atomic array with $N_{rad}=25$ particles in the radial direction and for different lattice constants $d$. In (a), the cooperative decay rate $\bar{\Gamma}_1^{(av.)}$ is used, whereas (b) is obtained with $\bar{\Gamma}_1^{(mean)}$. (c)-(e) Maximum emission rate multiplied by $(d/\lambda)^3$ versus radius or characteristic length of the sample $N_{rad}d/\lambda$ for different values of $d$. The dashed lines are obtained with $\bar{\Gamma}_1^{(av.)}$, whereas the markers correspond to (c) $\bar{\Gamma}_1^{(mean)}$, (d) $\bar{\Gamma}_1^{(n.n.)}$, and (e) $\bar{\Gamma}_1^{(pair)}(\vec{r}_2=(|\vec{r}_2|,0,0))$. In (e), an array of spacing $d=0.2\lambda$ is considered.} 
\end{figure}

The results shown in the main text are obtained using $\bar{\Gamma}_1^{(av.)}$. In Fig.~\ref{fig: 3D_differentmethods}, we present the dynamics and values of the superradiant peak obtained with the other forms of the cooperative decay rate. In particular, we demonstrate the time evolution of the decaying ensemble for $\bar{\Gamma}_1^{(av.)}$ in Fig.~\ref{fig: 3D_differentmethods}(a) and for $\bar{\Gamma}_1^{(mean)}$ in Fig.~\ref{fig: 3D_differentmethods}(b). One can see that the resulting dynamics, that is, the superradiant burst, the subradiant phase, and the subsequent decay to the ground state of the system, are identical in both cases. Additionally, we obtain an identical value of the emission peak per particle $-\dot{a}_{max}$ for all forms of $\bar{\Gamma}_1$. In Figs.~\ref{fig: 3D_differentmethods}(c) and (d), we plot $-\dot{a}_{max}$ as a function of lattice size and for three different lattice spacings. The overlap between the dashed lines (which show the results obtained with $\bar{\Gamma}_1^{(av.)}$ and presented in the main text) and the markers [which correspond to the values computed with $\bar{\Gamma}_1^{(mean)}$ in Fig.~\ref{fig: 3D_differentmethods}(c) and with $\bar{\Gamma}_1^{(n.n.)}$ Fig.~\ref{fig: 3D_differentmethods}(d)] demonstrates that both methods result in the same scaling of the peak with the optical depth of the system. Similarly, Fig.~\ref{fig: 3D_differentmethods}(e) depicts $-\dot{a}_{max}$ for a lattice of spacing $d=0.2\lambda$ and for the collective decay rate $\bar{\Gamma}_1^{(pair)}$ computed for different distances $|\vec{r}_2|$ between probe atoms. Again, almost identical values are obtained independently of $|\vec{r}_2|$. These results confirm both the linear scaling of the superradiant peak with the optical depth of the array $\mathcal{O}=N_{rad}/d^2$ and the slight oscillations arising from the interference between different ``shells" of the lattice.

\section{Two-dimensional array}
\label{appendix: 2D array}

\begin{figure}[t!]
\includegraphics[width=\linewidth]{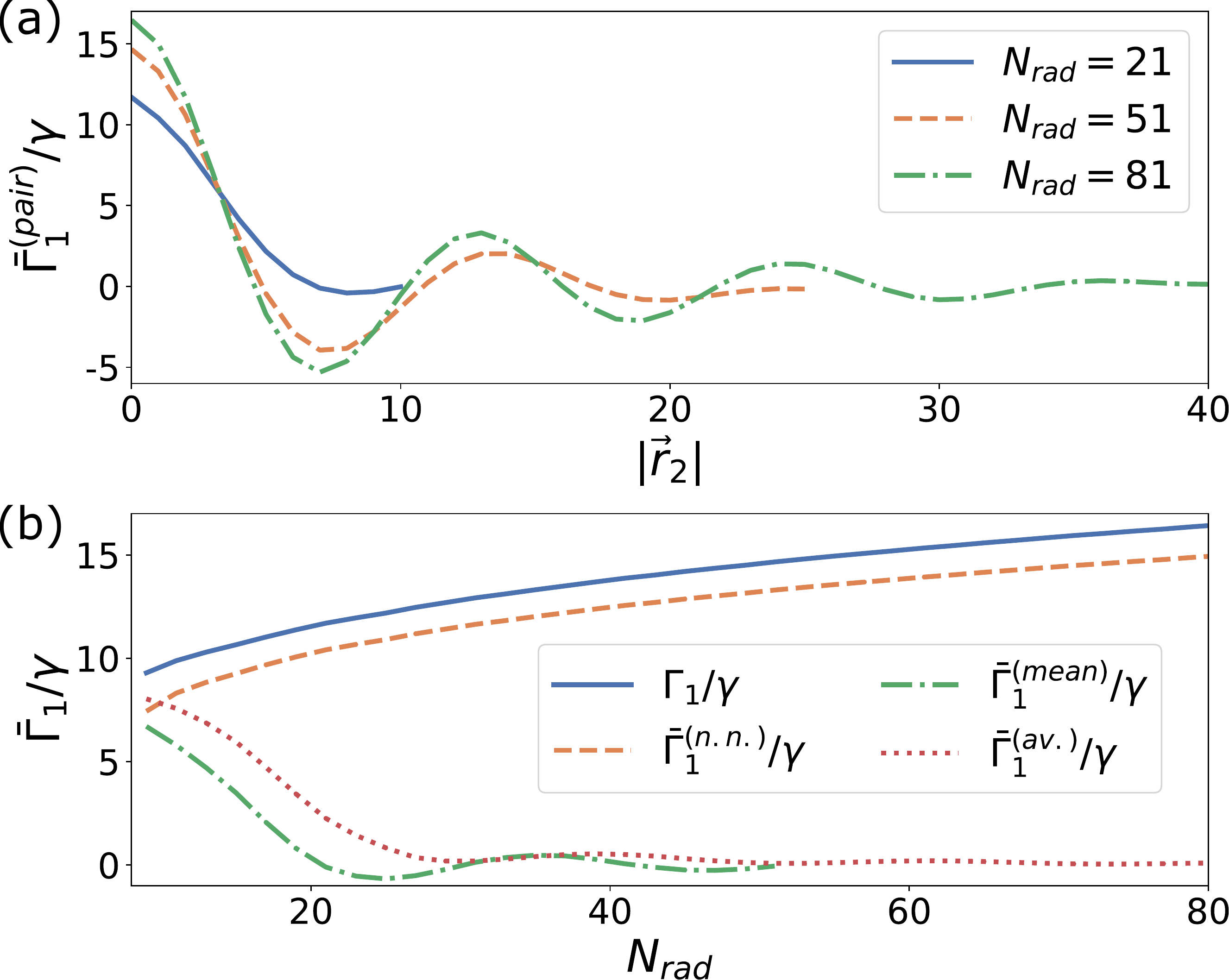}
\caption{\label{fig: Gammabar_2D_supplementary} \textbf{$\bar{\Gamma}_1$ in a two-dimensional array} (a) Plot of $\bar{\Gamma}_1^{(pair)}(\vec{r}=0,\vec{r}_2) /\gamma$ for a two-dimensional circular array of atoms with spacing $d=0.1\lambda$ and for different positions of the probe atom $\vec{r}_2=(|\vec{r}_2|,0)$. The different colors represent arrays of different sizes. (b) Plot of $\Gamma_1/\gamma$ (blue solid trace) and $\bar{\Gamma}_1/\gamma$ as a function of the number of atoms in the radial direction $N_{rad}$ for an array with $d=0.1\lambda$. The different colors and line styles correspond to the different ways of computing $\bar{\Gamma}_1$: $\bar{\Gamma}_1^{(n.n.)}$ as an orange dashed line, $\bar{\Gamma}_1^{(mean)}$ as a green dash-dotted line, and $\bar{\Gamma}_1^{(av.)}$ as a red dotted line. In both panels, we consider an initially inverted array $a=1$ and the corresponding $\Gamma$ given by Eq.~(\ref{eq: collectivedecayrates_1}).}
\end{figure}

Figure~\ref{fig: Gammabar_2D_supplementary}(a) shows $\bar{\Gamma}_1^{(pair)}(\vec{r}_1=0,\vec{r}_2)$ for circular two-dimensional arrays with different sizes $N_{rad}$ as a function of $\vec{r}_2=(|\vec{r}_2|,0)$. Again, in-phase and out-of-phase overlaps in the term $\sum_{\vec{x}} \tilde{D}^{ret} (\vec{x}) \tilde{D}^{*ret} (\vec{r}_2-\vec{x})$ result in maxima and minima of $\bar{\Gamma}_1$ and a subsequent oscillating behavior of $\bar{\Gamma}_1$ with the distance between probe atoms. Due to the absorbing nature of the two-dimensional Green's function, i.e., $\tilde{D}_{2D}^{ret}(\rho)$ decays with distance $\rho$, the oscillations are damped and the contribution of $\bar{\Gamma}_1$ to the two-atom cooperative decay rate becomes very small for probe atoms that lie far apart.

In Fig.~\ref{fig: Gammabar_2D_supplementary}(b), we compare the different ways of computing $\bar{\Gamma}_1$ for arrays of various sizes. As can be inferred from Fig.~\ref{fig: Gammabar_2D_supplementary}(a), $\bar{\Gamma}_1^{(n.n.)}(|\vec{r}_2|=d)$ is a growing function of the sample size and is close to $\Gamma_1=\bar{\Gamma}_1^{(n.n.)}(|\vec{r}_2|=0)$. The $\bar{\Gamma}_1^{(mean)}$ is obtained by averaging over all positions $\vec{r}_2$ present in the array and decays and oscillates with $N_{rad}$ due to the additional periods that emerge in $\bar{\Gamma}_1^{(pair)}(\vec{r}=0,\vec{r}_2)$ when the sample size is increased. Again, it follows a tren similar to $\bar{\Gamma}_1^{(av.)}$.

\begin{figure}[b]
\includegraphics[width=\linewidth]{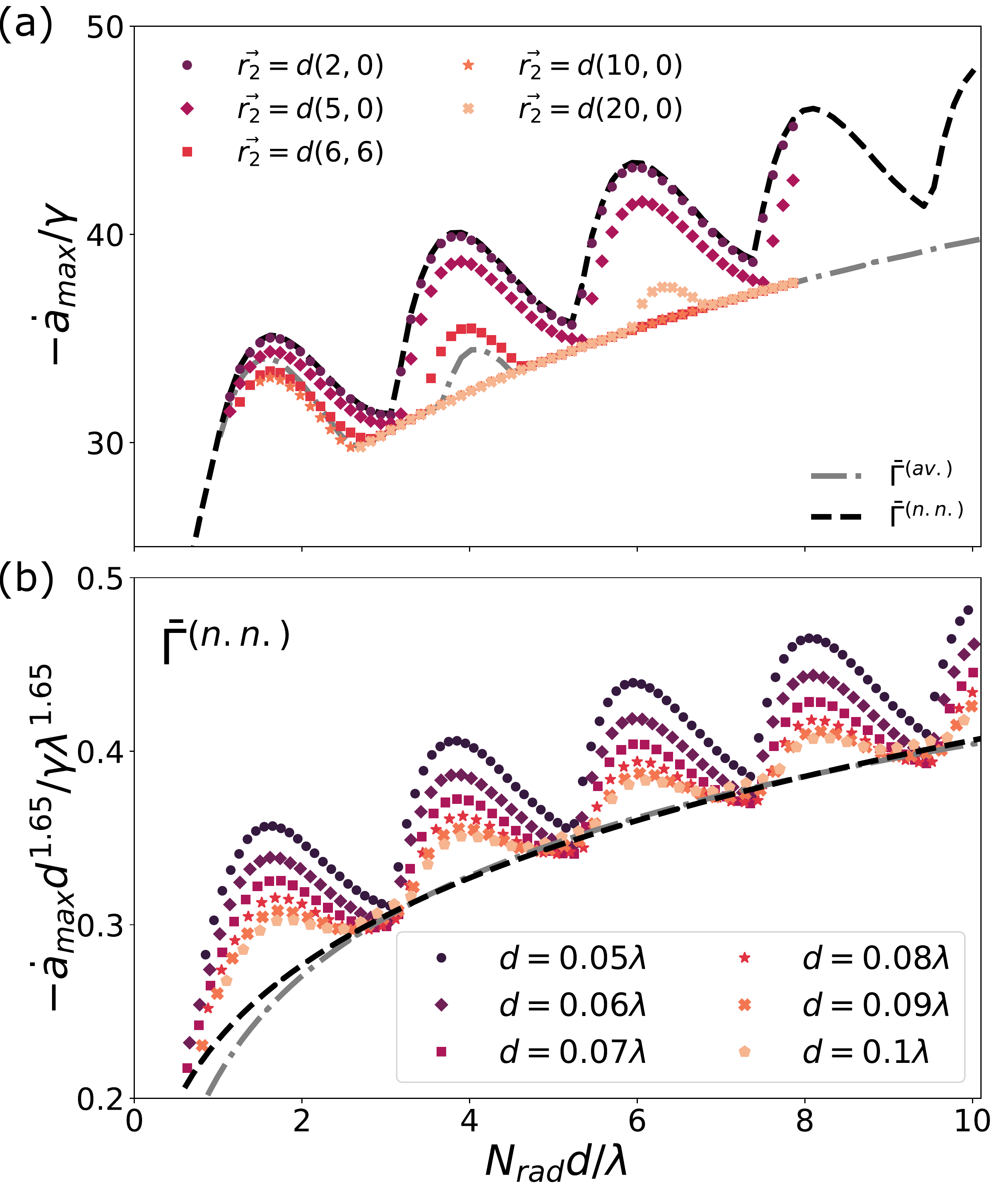}
\caption{\label{fig: peak_2D_supplementary} \textbf{Two-dimensional superradiant burst} (a) Maximum emission rate per particle $-\dot{a}_{max}$ resulting from $\bar{\Gamma}^{(pair)}$ for different probe atom pairs, represented by different colors. The black and gray dashed lines correspond to $-\dot{a}_{max}$ computed with $\bar{\Gamma}^{(n.n.)}$ and $\bar{\Gamma}^{(av.)}$, respectively. A lattice with spacing $d=0.06\lambda$ is considered. (b) Scaling of the minima of the superradiant peak resulting from $\bar{\Gamma}^{(n.n.)}$. The black curve corresponds to a fit of the form $-\dot{a}_{max} (d/\lambda)^{1.65} \propto (N_{rad}d/\lambda)^{0.22}$ and the gray trace to the functional form $-\dot{a}_{max} (d/\lambda)^{1.65} \propto \log (N_{rad}d/\lambda)$.}
\end{figure}

As opposed to the three-dimensional case, the value of the two-dimensional superradiant peak depends on the specific choice of $\bar{\Gamma}_1$. In the main text, we presented the results obtained using both $\bar{\Gamma}_1^{(n.n.)}$ and $\bar{\Gamma}_1^{(av.)}$. Note that the fact that $\bar{\Gamma}_1^{(n.n.)}>\bar{\Gamma}_1^{(av.)}$ results in larger values of the two-atom coherence and consequently of the superradiant peak for nearest neighbors, as shown in Fig.~\ref{fig: 2D_peak}(b). Also, the oscillations in $\Gamma_2=\bar{\Gamma}_2=\bar{\Gamma}_1^{(av.)} N\rho_{eg,ge}/a$ lead to an oscillatory behavior of the maximum emission rate $-\dot{a}_{max}$, which can be understood as an interference effect between different ``shells" of the array. In Fig.~\ref{fig: peak_2D_supplementary}(a), we complement the results reported in the main text with the maximum emission rate for probe atoms separated by different distances, obtained with $\bar{\Gamma}_1^{(pair)}$. One can see that the resulting $-\dot{a}_{max}$ is contained within the values retrieved from $\bar{\Gamma}_1^{(n.n.)}$ and $\bar{\Gamma}_1^{(av.)}$ for all distances $|\vec{r}_2|$ between the probe atoms. This suggests that the scaling of the superradiant peak is also contained within the values predicted using $\bar{\Gamma}_1^{(n.n.)}$ and $\bar{\Gamma}_1^{(av.)}$.

Finally, Fig.~\ref{fig: peak_2D_supplementary}(b) shows the maximum decay rate $-\dot{a}_{max}$ obtained with $\bar{\Gamma}_1^{(n.n.)}$ for arrays with different sizes and spacings. The minima of $-\dot{a}_{max}$ can be fitted both by the power law $-\dot{a}_{max} (d/\lambda)^{1.65} \propto (N_{rad}d/\lambda)^{0.22}$ (black dashed trace) and by the logarithmic function $-\dot{a}_{max} (d/\lambda)^{1.65} \propto \log (N_{rad}d/\lambda)$ (grey dash-dotted trace), which matches the scaling obtained in Fig.~\ref{fig: 2D_peak}(c) with $\bar{\Gamma}_1^{(av.)}$.

\nocite{*}

\bibliography{apssamp}

\begin{thebibliography}{59}%
\makeatletter
\providecommand \@ifxundefined [1]{%
 \@ifx{#1\undefined}
}%
\providecommand \@ifnum [1]{%
 \ifnum #1\expandafter \@firstoftwo
 \else \expandafter \@secondoftwo
 \fi
}%
\providecommand \@ifx [1]{%
 \ifx #1\expandafter \@firstoftwo
 \else \expandafter \@secondoftwo
 \fi
}%
\providecommand \natexlab [1]{#1}%
\providecommand \enquote  [1]{``#1''}%
\providecommand \bibnamefont  [1]{#1}%
\providecommand \bibfnamefont [1]{#1}%
\providecommand \citenamefont [1]{#1}%
\providecommand \href@noop [0]{\@secondoftwo}%
\providecommand \href [0]{\begingroup \@sanitize@url \@href}%
\providecommand \@href[1]{\@@startlink{#1}\@@href}%
\providecommand \@@href[1]{\endgroup#1\@@endlink}%
\providecommand \@sanitize@url [0]{\catcode `\\12\catcode `\$12\catcode
  `\&12\catcode `\#12\catcode `\^12\catcode `\_12\catcode `\%12\relax}%
\providecommand \@@startlink[1]{}%
\providecommand \@@endlink[0]{}%
\providecommand \url  [0]{\begingroup\@sanitize@url \@url }%
\providecommand \@url [1]{\endgroup\@href {#1}{\urlprefix }}%
\providecommand \urlprefix  [0]{URL }%
\providecommand \Eprint [0]{\href }%
\providecommand \doibase [0]{https://doi.org/}%
\providecommand \selectlanguage [0]{\@gobble}%
\providecommand \bibinfo  [0]{\@secondoftwo}%
\providecommand \bibfield  [0]{\@secondoftwo}%
\providecommand \translation [1]{[#1]}%
\providecommand \BibitemOpen [0]{}%
\providecommand \bibitemStop [0]{}%
\providecommand \bibitemNoStop [0]{.\EOS\space}%
\providecommand \EOS [0]{\spacefactor3000\relax}%
\providecommand \BibitemShut  [1]{\csname bibitem#1\endcsname}%
\let\auto@bib@innerbib\@empty
\bibitem [{\citenamefont {Dicke}(1954)}]{intro_Dicke}%
  \BibitemOpen
  \bibfield  {author} {\bibinfo {author} {\bibfnamefont {R.~H.}\ \bibnamefont
  {Dicke}},\ }\bibfield  {title} {\bibinfo {title} {Coherence in spontaneous
  radiation processes},\ }\href {https://doi.org/10.1103/PhysRev.93.99}
  {\bibfield  {journal} {\bibinfo  {journal} {Phys. Rev.}\ }\textbf {\bibinfo
  {volume} {93}},\ \bibinfo {pages} {99} (\bibinfo {year} {1954})}\BibitemShut
  {NoStop}%
\bibitem [{\citenamefont {Lehmberg}(1970)}]{intro_Lehmberg}%
  \BibitemOpen
  \bibfield  {author} {\bibinfo {author} {\bibfnamefont {R.~H.}\ \bibnamefont
  {Lehmberg}},\ }\bibfield  {title} {\bibinfo {title} {Radiation from an
  $n$-atom system. i. general formalism},\ }\href
  {https://doi.org/10.1103/PhysRevA.2.883} {\bibfield  {journal} {\bibinfo
  {journal} {Phys. Rev. A}\ }\textbf {\bibinfo {volume} {2}},\ \bibinfo {pages}
  {883} (\bibinfo {year} {1970})}\BibitemShut {NoStop}%
\bibitem [{\citenamefont {Gross}\ and\ \citenamefont
  {Haroche}(1982)}]{intro_gross}%
  \BibitemOpen
  \bibfield  {author} {\bibinfo {author} {\bibfnamefont {M.}~\bibnamefont
  {Gross}}\ and\ \bibinfo {author} {\bibfnamefont {S.}~\bibnamefont
  {Haroche}},\ }\href@noop {} {\bibfield  {journal} {\bibinfo  {journal} {Phys.
  Rep.}\ }\textbf {\bibinfo {volume} {93}},\ \bibinfo {pages} {301} (\bibinfo
  {year} {1982})}\BibitemShut {NoStop}%
\bibitem [{\citenamefont {Feher}\ \emph {et~al.}(1958)\citenamefont {Feher},
  \citenamefont {Gordon}, \citenamefont {Buehler}, \citenamefont {Gere},\ and\
  \citenamefont {Thurmond}}]{Experiment_1}%
  \BibitemOpen
  \bibfield  {author} {\bibinfo {author} {\bibfnamefont {G.}~\bibnamefont
  {Feher}}, \bibinfo {author} {\bibfnamefont {J.~P.}\ \bibnamefont {Gordon}},
  \bibinfo {author} {\bibfnamefont {E.}~\bibnamefont {Buehler}}, \bibinfo
  {author} {\bibfnamefont {E.~A.}\ \bibnamefont {Gere}},\ and\ \bibinfo
  {author} {\bibfnamefont {C.~D.}\ \bibnamefont {Thurmond}},\ }\bibfield
  {title} {\bibinfo {title} {Spontaneous emission of radiation from an electron
  spin system},\ }\href {https://doi.org/10.1103/PhysRev.109.221} {\bibfield
  {journal} {\bibinfo  {journal} {Phys. Rev.}\ }\textbf {\bibinfo {volume}
  {109}},\ \bibinfo {pages} {221} (\bibinfo {year} {1958})}\BibitemShut
  {NoStop}%
\bibitem [{\citenamefont {Skribanowitz}\ \emph {et~al.}(1973)\citenamefont
  {Skribanowitz}, \citenamefont {Herman}, \citenamefont {MacGillivray},\ and\
  \citenamefont {Feld}}]{Experiment_2}%
  \BibitemOpen
  \bibfield  {author} {\bibinfo {author} {\bibfnamefont {N.}~\bibnamefont
  {Skribanowitz}}, \bibinfo {author} {\bibfnamefont {I.~P.}\ \bibnamefont
  {Herman}}, \bibinfo {author} {\bibfnamefont {J.~C.}\ \bibnamefont
  {MacGillivray}},\ and\ \bibinfo {author} {\bibfnamefont {M.~S.}\ \bibnamefont
  {Feld}},\ }\bibfield  {title} {\bibinfo {title} {Observation of dicke
  superradiance in optically pumped hf gas},\ }\href
  {https://doi.org/10.1103/PhysRevLett.30.309} {\bibfield  {journal} {\bibinfo
  {journal} {Phys. Rev. Lett.}\ }\textbf {\bibinfo {volume} {30}},\ \bibinfo
  {pages} {309} (\bibinfo {year} {1973})}\BibitemShut {NoStop}%
\bibitem [{\citenamefont {Sz\"oke}\ and\ \citenamefont
  {Meiboom}(1959)}]{Experiment_thermal}%
  \BibitemOpen
  \bibfield  {author} {\bibinfo {author} {\bibfnamefont {A.}~\bibnamefont
  {Sz\"oke}}\ and\ \bibinfo {author} {\bibfnamefont {S.}~\bibnamefont
  {Meiboom}},\ }\bibfield  {title} {\bibinfo {title} {Radiation damping in
  nuclear magnetic resonance},\ }\href
  {https://doi.org/10.1103/PhysRev.113.585} {\bibfield  {journal} {\bibinfo
  {journal} {Phys. Rev.}\ }\textbf {\bibinfo {volume} {113}},\ \bibinfo {pages}
  {585} (\bibinfo {year} {1959})}\BibitemShut {NoStop}%
\bibitem [{\citenamefont {Inouye}\ \emph {et~al.}(1999)\citenamefont {Inouye},
  \citenamefont {Chikkatur}, \citenamefont {Stamper-Kurn}, \citenamefont
  {Stenger}, \citenamefont {Pritchard},\ and\ \citenamefont
  {Ketterle}}]{Experiment_3}%
  \BibitemOpen
  \bibfield  {author} {\bibinfo {author} {\bibfnamefont {S.}~\bibnamefont
  {Inouye}}, \bibinfo {author} {\bibfnamefont {A.~P.}\ \bibnamefont
  {Chikkatur}}, \bibinfo {author} {\bibfnamefont {D.~M.}\ \bibnamefont
  {Stamper-Kurn}}, \bibinfo {author} {\bibfnamefont {J.}~\bibnamefont
  {Stenger}}, \bibinfo {author} {\bibfnamefont {D.~E.}\ \bibnamefont
  {Pritchard}},\ and\ \bibinfo {author} {\bibfnamefont {W.}~\bibnamefont
  {Ketterle}},\ }\bibfield  {title} {\bibinfo {title} {Superradiant rayleigh
  scattering from a bose-einstein condensate},\ }\href
  {https://doi.org/10.1126/science.285.5427.571} {\bibfield  {journal}
  {\bibinfo  {journal} {Science}\ }\textbf {\bibinfo {volume} {285}},\ \bibinfo
  {pages} {571} (\bibinfo {year} {1999})}\BibitemShut {NoStop}%
\bibitem [{\citenamefont {Baumann}\ \emph {et~al.}(2010)\citenamefont
  {Baumann}, \citenamefont {Guerlin}, \citenamefont {Brennecke},\ and\
  \citenamefont {Esslinger}}]{Bose_cond_2}%
  \BibitemOpen
  \bibfield  {author} {\bibinfo {author} {\bibfnamefont {K.}~\bibnamefont
  {Baumann}}, \bibinfo {author} {\bibfnamefont {C.}~\bibnamefont {Guerlin}},
  \bibinfo {author} {\bibfnamefont {F.}~\bibnamefont {Brennecke}},\ and\
  \bibinfo {author} {\bibfnamefont {T.}~\bibnamefont {Esslinger}},\ }\bibfield
  {title} {\bibinfo {title} {Dicke quantum phase transition with a superfluid
  gas in an optical cavity},\ }\href {https://doi.org/10.1038/nature09009}
  {\bibfield  {journal} {\bibinfo  {journal} {Nature}\ }\textbf {\bibinfo
  {volume} {464}},\ \bibinfo {pages} {1301} (\bibinfo {year}
  {2010})}\BibitemShut {NoStop}%
\bibitem [{\citenamefont {Wang}\ \emph {et~al.}(2007)\citenamefont {Wang},
  \citenamefont {Yelin}, \citenamefont {C\^ot\'e}, \citenamefont {Eyler},
  \citenamefont {Farooqi}, \citenamefont {Gould}, \citenamefont
  {Ko\ifmmode~\check{s}\else \v{s}\fi{}trun}, \citenamefont {Tong},\ and\
  \citenamefont {Vrinceanu}}]{Experiment_4}%
  \BibitemOpen
  \bibfield  {author} {\bibinfo {author} {\bibfnamefont {T.}~\bibnamefont
  {Wang}}, \bibinfo {author} {\bibfnamefont {S.~F.}\ \bibnamefont {Yelin}},
  \bibinfo {author} {\bibfnamefont {R.}~\bibnamefont {C\^ot\'e}}, \bibinfo
  {author} {\bibfnamefont {E.~E.}\ \bibnamefont {Eyler}}, \bibinfo {author}
  {\bibfnamefont {S.~M.}\ \bibnamefont {Farooqi}}, \bibinfo {author}
  {\bibfnamefont {P.~L.}\ \bibnamefont {Gould}}, \bibinfo {author}
  {\bibfnamefont {M.}~\bibnamefont {Ko\ifmmode~\check{s}\else \v{s}\fi{}trun}},
  \bibinfo {author} {\bibfnamefont {D.}~\bibnamefont {Tong}},\ and\ \bibinfo
  {author} {\bibfnamefont {D.}~\bibnamefont {Vrinceanu}},\ }\bibfield  {title}
  {\bibinfo {title} {Superradiance in ultracold rydberg gases},\ }\href
  {https://doi.org/10.1103/PhysRevA.75.033802} {\bibfield  {journal} {\bibinfo
  {journal} {Phys. Rev. A}\ }\textbf {\bibinfo {volume} {75}},\ \bibinfo
  {pages} {033802} (\bibinfo {year} {2007})}\BibitemShut {NoStop}%
\bibitem [{\citenamefont {Kaluzny}\ \emph {et~al.}(1983)\citenamefont
  {Kaluzny}, \citenamefont {Goy}, \citenamefont {Gross}, \citenamefont
  {Raimond},\ and\ \citenamefont {Haroche}}]{Rydberg_2}%
  \BibitemOpen
  \bibfield  {author} {\bibinfo {author} {\bibfnamefont {Y.}~\bibnamefont
  {Kaluzny}}, \bibinfo {author} {\bibfnamefont {P.}~\bibnamefont {Goy}},
  \bibinfo {author} {\bibfnamefont {M.}~\bibnamefont {Gross}}, \bibinfo
  {author} {\bibfnamefont {J.~M.}\ \bibnamefont {Raimond}},\ and\ \bibinfo
  {author} {\bibfnamefont {S.}~\bibnamefont {Haroche}},\ }\bibfield  {title}
  {\bibinfo {title} {Observation of self-induced rabi oscillations in two-level
  atoms excited inside a resonant cavity: The ringing regime of
  superradiance},\ }\href {https://doi.org/10.1103/PhysRevLett.51.1175}
  {\bibfield  {journal} {\bibinfo  {journal} {Phys. Rev. Lett.}\ }\textbf
  {\bibinfo {volume} {51}},\ \bibinfo {pages} {1175} (\bibinfo {year}
  {1983})}\BibitemShut {NoStop}%
\bibitem [{\citenamefont {Grimes}\ \emph {et~al.}(2017)\citenamefont {Grimes},
  \citenamefont {Coy}, \citenamefont {Barnum}, \citenamefont {Zhou},
  \citenamefont {Yelin},\ and\ \citenamefont {Field}}]{Experiment_Rydberg}%
  \BibitemOpen
  \bibfield  {author} {\bibinfo {author} {\bibfnamefont {D.~D.}\ \bibnamefont
  {Grimes}}, \bibinfo {author} {\bibfnamefont {S.~L.}\ \bibnamefont {Coy}},
  \bibinfo {author} {\bibfnamefont {T.~J.}\ \bibnamefont {Barnum}}, \bibinfo
  {author} {\bibfnamefont {Y.}~\bibnamefont {Zhou}}, \bibinfo {author}
  {\bibfnamefont {S.~F.}\ \bibnamefont {Yelin}},\ and\ \bibinfo {author}
  {\bibfnamefont {R.~W.}\ \bibnamefont {Field}},\ }\bibfield  {title} {\bibinfo
  {title} {Direct single-shot observation of millimeter-wave superradiance in
  rydberg-rydberg transitions},\ }\href
  {https://doi.org/10.1103/PhysRevA.95.043818} {\bibfield  {journal} {\bibinfo
  {journal} {Phys. Rev. A}\ }\textbf {\bibinfo {volume} {95}},\ \bibinfo
  {pages} {043818} (\bibinfo {year} {2017})}\BibitemShut {NoStop}%
\bibitem [{\citenamefont {Bienaim\'e}\ \emph {et~al.}(2012)\citenamefont
  {Bienaim\'e}, \citenamefont {Piovella},\ and\ \citenamefont
  {Kaiser}}]{single_excitation_1}%
  \BibitemOpen
  \bibfield  {author} {\bibinfo {author} {\bibfnamefont {T.}~\bibnamefont
  {Bienaim\'e}}, \bibinfo {author} {\bibfnamefont {N.}~\bibnamefont
  {Piovella}},\ and\ \bibinfo {author} {\bibfnamefont {R.}~\bibnamefont
  {Kaiser}},\ }\bibfield  {title} {\bibinfo {title} {Controlled dicke
  subradiance from a large cloud of two-level systems},\ }\href
  {https://doi.org/10.1103/PhysRevLett.108.123602} {\bibfield  {journal}
  {\bibinfo  {journal} {Phys. Rev. Lett.}\ }\textbf {\bibinfo {volume} {108}},\
  \bibinfo {pages} {123602} (\bibinfo {year} {2012})}\BibitemShut {NoStop}%
\bibitem [{\citenamefont {Scully}\ \emph {et~al.}(2006)\citenamefont {Scully},
  \citenamefont {Fry}, \citenamefont {Ooi},\ and\ \citenamefont
  {W\'odkiewicz}}]{single_excitation_2}%
  \BibitemOpen
  \bibfield  {author} {\bibinfo {author} {\bibfnamefont {M.~O.}\ \bibnamefont
  {Scully}}, \bibinfo {author} {\bibfnamefont {E.~S.}\ \bibnamefont {Fry}},
  \bibinfo {author} {\bibfnamefont {C.~H.~R.}\ \bibnamefont {Ooi}},\ and\
  \bibinfo {author} {\bibfnamefont {K.}~\bibnamefont {W\'odkiewicz}},\
  }\bibfield  {title} {\bibinfo {title} {Directed spontaneous emission from an
  extended ensemble of $n$ atoms: Timing is everything},\ }\href
  {https://doi.org/10.1103/PhysRevLett.96.010501} {\bibfield  {journal}
  {\bibinfo  {journal} {Phys. Rev. Lett.}\ }\textbf {\bibinfo {volume} {96}},\
  \bibinfo {pages} {010501} (\bibinfo {year} {2006})}\BibitemShut {NoStop}%
\bibitem [{\citenamefont {Svidzinsky}\ and\ \citenamefont
  {Chang}(2008)}]{single_excitation_3}%
  \BibitemOpen
  \bibfield  {author} {\bibinfo {author} {\bibfnamefont {A.}~\bibnamefont
  {Svidzinsky}}\ and\ \bibinfo {author} {\bibfnamefont {J.-T.}\ \bibnamefont
  {Chang}},\ }\bibfield  {title} {\bibinfo {title} {Cooperative spontaneous
  emission as a many-body eigenvalue problem},\ }\href
  {https://doi.org/10.1103/PhysRevA.77.043833} {\bibfield  {journal} {\bibinfo
  {journal} {Phys. Rev. A}\ }\textbf {\bibinfo {volume} {77}},\ \bibinfo
  {pages} {043833} (\bibinfo {year} {2008})}\BibitemShut {NoStop}%
\bibitem [{\citenamefont {Svidzinsky}\ \emph {et~al.}(2010)\citenamefont
  {Svidzinsky}, \citenamefont {Chang},\ and\ \citenamefont
  {Scully}}]{single_excitation_4}%
  \BibitemOpen
  \bibfield  {author} {\bibinfo {author} {\bibfnamefont {A.~A.}\ \bibnamefont
  {Svidzinsky}}, \bibinfo {author} {\bibfnamefont {J.-T.}\ \bibnamefont
  {Chang}},\ and\ \bibinfo {author} {\bibfnamefont {M.~O.}\ \bibnamefont
  {Scully}},\ }\bibfield  {title} {\bibinfo {title} {Cooperative spontaneous
  emission of $n$ atoms: Many-body eigenstates, the effect of virtual lamb
  shift processes, and analogy with radiation of $n$ classical oscillators},\
  }\href {https://doi.org/10.1103/PhysRevA.81.053821} {\bibfield  {journal}
  {\bibinfo  {journal} {Phys. Rev. A}\ }\textbf {\bibinfo {volume} {81}},\
  \bibinfo {pages} {053821} (\bibinfo {year} {2010})}\BibitemShut {NoStop}%
\bibitem [{\citenamefont {Kong}\ and\ \citenamefont
  {P\'alffy}(2017)}]{single_excitatin_5}%
  \BibitemOpen
  \bibfield  {author} {\bibinfo {author} {\bibfnamefont {X.}~\bibnamefont
  {Kong}}\ and\ \bibinfo {author} {\bibfnamefont {A.}~\bibnamefont
  {P\'alffy}},\ }\bibfield  {title} {\bibinfo {title} {Collective radiation
  spectrum for ensembles with zeeman splitting in single-photon
  superradiance},\ }\href {https://doi.org/10.1103/PhysRevA.96.033819}
  {\bibfield  {journal} {\bibinfo  {journal} {Phys. Rev. A}\ }\textbf {\bibinfo
  {volume} {96}},\ \bibinfo {pages} {033819} (\bibinfo {year}
  {2017})}\BibitemShut {NoStop}%
\bibitem [{\citenamefont {Cottier}\ \emph {et~al.}(2018)\citenamefont
  {Cottier}, \citenamefont {Kaiser},\ and\ \citenamefont
  {Bachelard}}]{single_excitation_6}%
  \BibitemOpen
  \bibfield  {author} {\bibinfo {author} {\bibfnamefont {F.}~\bibnamefont
  {Cottier}}, \bibinfo {author} {\bibfnamefont {R.}~\bibnamefont {Kaiser}},\
  and\ \bibinfo {author} {\bibfnamefont {R.}~\bibnamefont {Bachelard}},\
  }\bibfield  {title} {\bibinfo {title} {Role of disorder in super- and
  subradiance of cold atomic clouds},\ }\href
  {https://doi.org/10.1103/PhysRevA.98.013622} {\bibfield  {journal} {\bibinfo
  {journal} {Phys. Rev. A}\ }\textbf {\bibinfo {volume} {98}},\ \bibinfo
  {pages} {013622} (\bibinfo {year} {2018})}\BibitemShut {NoStop}%
\bibitem [{\citenamefont {Clemens}\ \emph {et~al.}(2003)\citenamefont
  {Clemens}, \citenamefont {Horvath}, \citenamefont {Sanders},\ and\
  \citenamefont {Carmichael}}]{few_atoms_charmichael}%
  \BibitemOpen
  \bibfield  {author} {\bibinfo {author} {\bibfnamefont {J.~P.}\ \bibnamefont
  {Clemens}}, \bibinfo {author} {\bibfnamefont {L.}~\bibnamefont {Horvath}},
  \bibinfo {author} {\bibfnamefont {B.~C.}\ \bibnamefont {Sanders}},\ and\
  \bibinfo {author} {\bibfnamefont {H.~J.}\ \bibnamefont {Carmichael}},\
  }\bibfield  {title} {\bibinfo {title} {Collective spontaneous emission from a
  line of atoms},\ }\href {https://doi.org/10.1103/PhysRevA.68.023809}
  {\bibfield  {journal} {\bibinfo  {journal} {Phys. Rev. A}\ }\textbf {\bibinfo
  {volume} {68}},\ \bibinfo {pages} {023809} (\bibinfo {year}
  {2003})}\BibitemShut {NoStop}%
\bibitem [{\citenamefont {Masson}\ \emph {et~al.}(2020)\citenamefont {Masson},
  \citenamefont {Ferrier-Barbut}, \citenamefont {Orozco}, \citenamefont
  {Browaeys},\ and\ \citenamefont {Asenjo-Garcia}}]{few_atoms_Anna}%
  \BibitemOpen
  \bibfield  {author} {\bibinfo {author} {\bibfnamefont {S.~J.}\ \bibnamefont
  {Masson}}, \bibinfo {author} {\bibfnamefont {I.}~\bibnamefont
  {Ferrier-Barbut}}, \bibinfo {author} {\bibfnamefont {L.~A.}\ \bibnamefont
  {Orozco}}, \bibinfo {author} {\bibfnamefont {A.}~\bibnamefont {Browaeys}},\
  and\ \bibinfo {author} {\bibfnamefont {A.}~\bibnamefont {Asenjo-Garcia}},\
  }\bibfield  {title} {\bibinfo {title} {Many-body signatures of collective
  decay in atomic chains},\ }\href
  {https://doi.org/10.1103/PhysRevLett.125.263601} {\bibfield  {journal}
  {\bibinfo  {journal} {Phys. Rev. Lett.}\ }\textbf {\bibinfo {volume} {125}},\
  \bibinfo {pages} {263601} (\bibinfo {year} {2020})}\BibitemShut {NoStop}%
\bibitem [{\citenamefont {Carmichael}\ and\ \citenamefont
  {Kim}(2000)}]{charmichael_2}%
  \BibitemOpen
  \bibfield  {author} {\bibinfo {author} {\bibfnamefont {H.}~\bibnamefont
  {Carmichael}}\ and\ \bibinfo {author} {\bibfnamefont {K.}~\bibnamefont
  {Kim}},\ }\bibfield  {title} {\bibinfo {title} {A quantum trajectory
  unraveling of the superradiance master equation.},\ }\href
  {https://doi.org/https://doi.org/10.1016/S0030-4018(99)00694-X} {\bibfield
  {journal} {\bibinfo  {journal} {Optics Communications}\ }\textbf {\bibinfo
  {volume} {179}},\ \bibinfo {pages} {417} (\bibinfo {year}
  {2000})}\BibitemShut {NoStop}%
\bibitem [{\citenamefont {Rui}\ \emph {et~al.}(2020)\citenamefont {Rui},
  \citenamefont {Wei}, \citenamefont {Rubio-Abadal}, \citenamefont {Hollerith},
  \citenamefont {Zeiher}, \citenamefont {Stamper-Kurn}, \citenamefont {Gross},\
  and\ \citenamefont {Bloch}}]{OpticalLAttice_1}%
  \BibitemOpen
  \bibfield  {author} {\bibinfo {author} {\bibfnamefont {J.}~\bibnamefont
  {Rui}}, \bibinfo {author} {\bibfnamefont {D.}~\bibnamefont {Wei}}, \bibinfo
  {author} {\bibfnamefont {A.}~\bibnamefont {Rubio-Abadal}}, \bibinfo {author}
  {\bibfnamefont {S.}~\bibnamefont {Hollerith}}, \bibinfo {author}
  {\bibfnamefont {J.}~\bibnamefont {Zeiher}}, \bibinfo {author} {\bibfnamefont
  {D.~M.}\ \bibnamefont {Stamper-Kurn}}, \bibinfo {author} {\bibfnamefont
  {C.}~\bibnamefont {Gross}},\ and\ \bibinfo {author} {\bibfnamefont
  {I.}~\bibnamefont {Bloch}},\ }\bibfield  {title} {\bibinfo {title} {A
  subradiant optical mirror formed by a single structured atomic layer},\
  }\href@noop {} {\bibfield  {journal} {\bibinfo  {journal} {Nature}\ }\textbf
  {\bibinfo {volume} {583}},\ \bibinfo {pages} {369374} (\bibinfo {year}
  {2020})}\BibitemShut {NoStop}%
\bibitem [{\citenamefont {Bloch}(2005)}]{OpticalLAttice_2}%
  \BibitemOpen
  \bibfield  {author} {\bibinfo {author} {\bibfnamefont {I.}~\bibnamefont
  {Bloch}},\ }\bibfield  {title} {\bibinfo {title} {Ultracold quantum gases in
  optical lattices},\ }\href@noop {} {\bibfield  {journal} {\bibinfo  {journal}
  {Nature Physics}\ }\textbf {\bibinfo {volume} {1}},\ \bibinfo {pages} {23}
  (\bibinfo {year} {2005})}\BibitemShut {NoStop}%
\bibitem [{\citenamefont {Labuhn}\ \emph {et~al.}(2016)\citenamefont {Labuhn},
  \citenamefont {Barredo}, \citenamefont {Ravets}, \citenamefont
  {de~Léséleuc}, \citenamefont {Macrì}, \citenamefont {Lahaye},\ and\
  \citenamefont {Browaeys}}]{OpticalLAttice_3}%
  \BibitemOpen
  \bibfield  {author} {\bibinfo {author} {\bibfnamefont {H.}~\bibnamefont
  {Labuhn}}, \bibinfo {author} {\bibfnamefont {D.}~\bibnamefont {Barredo}},
  \bibinfo {author} {\bibfnamefont {S.}~\bibnamefont {Ravets}}, \bibinfo
  {author} {\bibfnamefont {S.}~\bibnamefont {de~Léséleuc}}, \bibinfo {author}
  {\bibfnamefont {T.}~\bibnamefont {Macrì}}, \bibinfo {author} {\bibfnamefont
  {T.}~\bibnamefont {Lahaye}},\ and\ \bibinfo {author} {\bibfnamefont
  {A.}~\bibnamefont {Browaeys}},\ }\bibfield  {title} {\bibinfo {title}
  {Tunable two-dimensional arrays of single rydberg atoms for realizing quantum
  ising models},\ }\href@noop {} {\bibfield  {journal} {\bibinfo  {journal}
  {Nature}\ }\textbf {\bibinfo {volume} {534}},\ \bibinfo {pages} {667670}
  (\bibinfo {year} {2016})}\BibitemShut {NoStop}%
\bibitem [{\citenamefont {Endres}\ \emph {et~al.}(2016)\citenamefont {Endres},
  \citenamefont {Bernien}, \citenamefont {Keesling}, \citenamefont {Levine},
  \citenamefont {Anschuetz}, \citenamefont {Krajenbrink}, \citenamefont
  {Senko}, \citenamefont {Vuletic}, \citenamefont {Greiner},\ and\
  \citenamefont {Lukin}}]{Tweezer_1}%
  \BibitemOpen
  \bibfield  {author} {\bibinfo {author} {\bibfnamefont {M.}~\bibnamefont
  {Endres}}, \bibinfo {author} {\bibfnamefont {H.}~\bibnamefont {Bernien}},
  \bibinfo {author} {\bibfnamefont {A.}~\bibnamefont {Keesling}}, \bibinfo
  {author} {\bibfnamefont {H.}~\bibnamefont {Levine}}, \bibinfo {author}
  {\bibfnamefont {E.~R.}\ \bibnamefont {Anschuetz}}, \bibinfo {author}
  {\bibfnamefont {A.}~\bibnamefont {Krajenbrink}}, \bibinfo {author}
  {\bibfnamefont {C.}~\bibnamefont {Senko}}, \bibinfo {author} {\bibfnamefont
  {V.}~\bibnamefont {Vuletic}}, \bibinfo {author} {\bibfnamefont
  {M.}~\bibnamefont {Greiner}},\ and\ \bibinfo {author} {\bibfnamefont {M.~D.}\
  \bibnamefont {Lukin}},\ }\bibfield  {title} {\bibinfo {title} {Atom-by-atom
  assembly of defect-free one-dimensional cold atom arrays},\ }\href
  {https://doi.org/10.1126/science.aah3752} {\bibfield  {journal} {\bibinfo
  {journal} {Science}\ }\textbf {\bibinfo {volume} {354}},\ \bibinfo {pages}
  {1024} (\bibinfo {year} {2016})}\BibitemShut {NoStop}%
\bibitem [{\citenamefont {Barredo}\ \emph {et~al.}(2016)\citenamefont
  {Barredo}, \citenamefont {de~L{\'e}s{\'e}leuc}, \citenamefont {Lienhard},
  \citenamefont {Lahaye},\ and\ \citenamefont {Browaeys}}]{Tweezer_2}%
  \BibitemOpen
  \bibfield  {author} {\bibinfo {author} {\bibfnamefont {D.}~\bibnamefont
  {Barredo}}, \bibinfo {author} {\bibfnamefont {S.}~\bibnamefont
  {de~L{\'e}s{\'e}leuc}}, \bibinfo {author} {\bibfnamefont {V.}~\bibnamefont
  {Lienhard}}, \bibinfo {author} {\bibfnamefont {T.}~\bibnamefont {Lahaye}},\
  and\ \bibinfo {author} {\bibfnamefont {A.}~\bibnamefont {Browaeys}},\
  }\bibfield  {title} {\bibinfo {title} {An atom-by-atom assembler of
  defect-free arbitrary two-dimensional atomic arrays},\ }\href
  {https://doi.org/10.1126/science.aah3778} {\bibfield  {journal} {\bibinfo
  {journal} {Science}\ }\textbf {\bibinfo {volume} {354}},\ \bibinfo {pages}
  {1021} (\bibinfo {year} {2016})}\BibitemShut {NoStop}%
\bibitem [{\citenamefont {Ohl~de Mello}\ \emph {et~al.}(2019)\citenamefont
  {Ohl~de Mello}, \citenamefont {Sch\"affner}, \citenamefont {Werkmann},
  \citenamefont {Preuschoff}, \citenamefont {Kohfahl}, \citenamefont
  {Schlosser},\ and\ \citenamefont {Birkl}}]{Tweezer_3}%
  \BibitemOpen
  \bibfield  {author} {\bibinfo {author} {\bibfnamefont {D.}~\bibnamefont
  {Ohl~de Mello}}, \bibinfo {author} {\bibfnamefont {D.}~\bibnamefont
  {Sch\"affner}}, \bibinfo {author} {\bibfnamefont {J.}~\bibnamefont
  {Werkmann}}, \bibinfo {author} {\bibfnamefont {T.}~\bibnamefont
  {Preuschoff}}, \bibinfo {author} {\bibfnamefont {L.}~\bibnamefont {Kohfahl}},
  \bibinfo {author} {\bibfnamefont {M.}~\bibnamefont {Schlosser}},\ and\
  \bibinfo {author} {\bibfnamefont {G.}~\bibnamefont {Birkl}},\ }\bibfield
  {title} {\bibinfo {title} {Defect-free assembly of 2d clusters of more than
  100 single-atom quantum systems},\ }\href
  {https://doi.org/10.1103/PhysRevLett.122.203601} {\bibfield  {journal}
  {\bibinfo  {journal} {Phys. Rev. Lett.}\ }\textbf {\bibinfo {volume} {122}},\
  \bibinfo {pages} {203601} (\bibinfo {year} {2019})}\BibitemShut {NoStop}%
\bibitem [{\citenamefont {Shahmoon}\ \emph {et~al.}(2017)\citenamefont
  {Shahmoon}, \citenamefont {Wild}, \citenamefont {Lukin},\ and\ \citenamefont
  {Yelin}}]{OneExcitation_Ephi}%
  \BibitemOpen
  \bibfield  {author} {\bibinfo {author} {\bibfnamefont {E.}~\bibnamefont
  {Shahmoon}}, \bibinfo {author} {\bibfnamefont {D.~S.}\ \bibnamefont {Wild}},
  \bibinfo {author} {\bibfnamefont {M.~D.}\ \bibnamefont {Lukin}},\ and\
  \bibinfo {author} {\bibfnamefont {S.~F.}\ \bibnamefont {Yelin}},\ }\bibfield
  {title} {\bibinfo {title} {Cooperative resonances in light scattering from
  two-dimensional atomic arrays},\ }\href
  {https://doi.org/10.1103/PhysRevLett.118.113601} {\bibfield  {journal}
  {\bibinfo  {journal} {Phys. Rev. Lett.}\ }\textbf {\bibinfo {volume} {118}},\
  \bibinfo {pages} {113601} (\bibinfo {year} {2017})}\BibitemShut {NoStop}%
\bibitem [{\citenamefont {Asenjo-Garcia}\ \emph {et~al.}(2017)\citenamefont
  {Asenjo-Garcia}, \citenamefont {Moreno-Cardoner}, \citenamefont {Albrecht},
  \citenamefont {Kimble},\ and\ \citenamefont {Chang}}]{OneExcitation_Ana}%
  \BibitemOpen
  \bibfield  {author} {\bibinfo {author} {\bibfnamefont {A.}~\bibnamefont
  {Asenjo-Garcia}}, \bibinfo {author} {\bibfnamefont {M.}~\bibnamefont
  {Moreno-Cardoner}}, \bibinfo {author} {\bibfnamefont {A.}~\bibnamefont
  {Albrecht}}, \bibinfo {author} {\bibfnamefont {H.~J.}\ \bibnamefont
  {Kimble}},\ and\ \bibinfo {author} {\bibfnamefont {D.~E.}\ \bibnamefont
  {Chang}},\ }\bibfield  {title} {\bibinfo {title} {Exponential improvement in
  photon storage fidelities using subradiance and ``selective radiance'' in
  atomic arrays},\ }\href {https://doi.org/10.1103/PhysRevX.7.031024}
  {\bibfield  {journal} {\bibinfo  {journal} {Phys. Rev. X}\ }\textbf {\bibinfo
  {volume} {7}},\ \bibinfo {pages} {031024} (\bibinfo {year}
  {2017})}\BibitemShut {NoStop}%
\bibitem [{\citenamefont {Henriet}\ \emph {et~al.}(2019)\citenamefont
  {Henriet}, \citenamefont {Douglas}, \citenamefont {Chang},\ and\
  \citenamefont {Albrecht}}]{OneExcitation_Chang}%
  \BibitemOpen
  \bibfield  {author} {\bibinfo {author} {\bibfnamefont {L.}~\bibnamefont
  {Henriet}}, \bibinfo {author} {\bibfnamefont {J.~S.}\ \bibnamefont
  {Douglas}}, \bibinfo {author} {\bibfnamefont {D.~E.}\ \bibnamefont {Chang}},\
  and\ \bibinfo {author} {\bibfnamefont {A.}~\bibnamefont {Albrecht}},\
  }\bibfield  {title} {\bibinfo {title} {Critical open-system dynamics in a
  one-dimensional optical-lattice clock},\ }\href
  {https://doi.org/10.1103/PhysRevA.99.023802} {\bibfield  {journal} {\bibinfo
  {journal} {Phys. Rev. A}\ }\textbf {\bibinfo {volume} {99}},\ \bibinfo
  {pages} {023802} (\bibinfo {year} {2019})}\BibitemShut {NoStop}%
\bibitem [{\citenamefont {Moreno-Cardoner}\ \emph {et~al.}(2019)\citenamefont
  {Moreno-Cardoner}, \citenamefont {Plankensteiner}, \citenamefont {Ostermann},
  \citenamefont {Chang},\ and\ \citenamefont {Ritsch}}]{one_exc_array_mariona}%
  \BibitemOpen
  \bibfield  {author} {\bibinfo {author} {\bibfnamefont {M.}~\bibnamefont
  {Moreno-Cardoner}}, \bibinfo {author} {\bibfnamefont {D.}~\bibnamefont
  {Plankensteiner}}, \bibinfo {author} {\bibfnamefont {L.}~\bibnamefont
  {Ostermann}}, \bibinfo {author} {\bibfnamefont {D.~E.}\ \bibnamefont
  {Chang}},\ and\ \bibinfo {author} {\bibfnamefont {H.}~\bibnamefont
  {Ritsch}},\ }\bibfield  {title} {\bibinfo {title} {Subradiance-enhanced
  excitation transfer between dipole-coupled nanorings of quantum emitters},\
  }\href {https://doi.org/10.1103/PhysRevA.100.023806} {\bibfield  {journal}
  {\bibinfo  {journal} {Phys. Rev. A}\ }\textbf {\bibinfo {volume} {100}},\
  \bibinfo {pages} {023806} (\bibinfo {year} {2019})}\BibitemShut {NoStop}%
\bibitem [{\citenamefont {Patti}\ \emph {et~al.}(2021)\citenamefont {Patti},
  \citenamefont {Wild}, \citenamefont {Shahmoon}, \citenamefont {Lukin},\ and\
  \citenamefont {Yelin}}]{one_exc_array_taylor}%
  \BibitemOpen
  \bibfield  {author} {\bibinfo {author} {\bibfnamefont {T.~L.}\ \bibnamefont
  {Patti}}, \bibinfo {author} {\bibfnamefont {D.~S.}\ \bibnamefont {Wild}},
  \bibinfo {author} {\bibfnamefont {E.}~\bibnamefont {Shahmoon}}, \bibinfo
  {author} {\bibfnamefont {M.~D.}\ \bibnamefont {Lukin}},\ and\ \bibinfo
  {author} {\bibfnamefont {S.~F.}\ \bibnamefont {Yelin}},\ }\bibfield  {title}
  {\bibinfo {title} {Controlling interactions between quantum emitters using
  atom arrays},\ }\href {https://doi.org/10.1103/PhysRevLett.126.223602}
  {\bibfield  {journal} {\bibinfo  {journal} {Phys. Rev. Lett.}\ }\textbf
  {\bibinfo {volume} {126}},\ \bibinfo {pages} {223602} (\bibinfo {year}
  {2021})}\BibitemShut {NoStop}%
\bibitem [{\citenamefont {Bettles}\ \emph {et~al.}(2016)\citenamefont
  {Bettles}, \citenamefont {Gardiner},\ and\ \citenamefont
  {Adams}}]{one_exc_array_adams}%
  \BibitemOpen
  \bibfield  {author} {\bibinfo {author} {\bibfnamefont {R.~J.}\ \bibnamefont
  {Bettles}}, \bibinfo {author} {\bibfnamefont {S.~A.}\ \bibnamefont
  {Gardiner}},\ and\ \bibinfo {author} {\bibfnamefont {C.~S.}\ \bibnamefont
  {Adams}},\ }\bibfield  {title} {\bibinfo {title} {Cooperative eigenmodes and
  scattering in one-dimensional atomic arrays},\ }\href
  {https://doi.org/10.1103/PhysRevA.94.043844} {\bibfield  {journal} {\bibinfo
  {journal} {Phys. Rev. A}\ }\textbf {\bibinfo {volume} {94}},\ \bibinfo
  {pages} {043844} (\bibinfo {year} {2016})}\BibitemShut {NoStop}%
\bibitem [{\citenamefont {Parmee}\ and\ \citenamefont
  {Ruostekoski}(2020)}]{one_excitation_roustekoski}%
  \BibitemOpen
  \bibfield  {author} {\bibinfo {author} {\bibfnamefont {C.~D.}\ \bibnamefont
  {Parmee}}\ and\ \bibinfo {author} {\bibfnamefont {J.}~\bibnamefont
  {Ruostekoski}},\ }\bibfield  {title} {\bibinfo {title} {Signatures of optical
  phase transitions in superradiant and subradiant atomic arrays},\ }\href
  {https://doi.org/10.1038/s42005-020-00476-1} {\bibfield  {journal} {\bibinfo
  {journal} {Communications Physics}\ }\textbf {\bibinfo {volume} {3}},\
  \bibinfo {pages} {205} (\bibinfo {year} {2020})}\BibitemShut {NoStop}%
\bibitem [{\citenamefont {Rubies-Bigorda}\ \emph {et~al.}(2022)\citenamefont
  {Rubies-Bigorda}, \citenamefont {Walther}, \citenamefont {Patti},\ and\
  \citenamefont {Yelin}}]{OneExcitation_me}%
  \BibitemOpen
  \bibfield  {author} {\bibinfo {author} {\bibfnamefont {O.}~\bibnamefont
  {Rubies-Bigorda}}, \bibinfo {author} {\bibfnamefont {V.}~\bibnamefont
  {Walther}}, \bibinfo {author} {\bibfnamefont {T.~L.}\ \bibnamefont {Patti}},\
  and\ \bibinfo {author} {\bibfnamefont {S.~F.}\ \bibnamefont {Yelin}},\
  }\bibfield  {title} {\bibinfo {title} {Photon control and coherent
  interactions via lattice dark states in atomic arrays},\ }\href
  {https://doi.org/10.1103/PhysRevResearch.4.013110} {\bibfield  {journal}
  {\bibinfo  {journal} {Phys. Rev. Research}\ }\textbf {\bibinfo {volume}
  {4}},\ \bibinfo {pages} {013110} (\bibinfo {year} {2022})}\BibitemShut
  {NoStop}%
\bibitem [{\citenamefont {Masson}\ and\ \citenamefont
  {Asenjo-Garcia}(2022)}]{superradiantburst_ana}%
  \BibitemOpen
  \bibfield  {author} {\bibinfo {author} {\bibfnamefont {S.~J.}\ \bibnamefont
  {Masson}}\ and\ \bibinfo {author} {\bibfnamefont {A.}~\bibnamefont
  {Asenjo-Garcia}},\ }\bibfield  {title} {\bibinfo {title} {Universality of
  dicke superradiance in arrays of quantum emitters},\ }\href
  {https://doi.org/10.1038/s41467-022-29805-4} {\bibfield  {journal} {\bibinfo
  {journal} {Nature Communications}\ }\textbf {\bibinfo {volume} {13}},\
  \bibinfo {pages} {2285} (\bibinfo {year} {2022})}\BibitemShut {NoStop}%
\bibitem [{\citenamefont {Sierra}\ \emph {et~al.}(2022)\citenamefont {Sierra},
  \citenamefont {Masson},\ and\ \citenamefont {Asenjo-Garcia}}]{Ana_super_new}%
  \BibitemOpen
  \bibfield  {author} {\bibinfo {author} {\bibfnamefont {E.}~\bibnamefont
  {Sierra}}, \bibinfo {author} {\bibfnamefont {S.~J.}\ \bibnamefont {Masson}},\
  and\ \bibinfo {author} {\bibfnamefont {A.}~\bibnamefont {Asenjo-Garcia}},\
  }\bibfield  {title} {\bibinfo {title} {Dicke superradiance in ordered
  lattices: Dimensionality matters},\ }\href
  {https://doi.org/10.1103/PhysRevResearch.4.023207} {\bibfield  {journal}
  {\bibinfo  {journal} {Phys. Rev. Research}\ }\textbf {\bibinfo {volume}
  {4}},\ \bibinfo {pages} {023207} (\bibinfo {year} {2022})}\BibitemShut
  {NoStop}%
\bibitem [{\citenamefont {Robicheaux}(2021)}]{robicheaux_superradiance}%
  \BibitemOpen
  \bibfield  {author} {\bibinfo {author} {\bibfnamefont {F.}~\bibnamefont
  {Robicheaux}},\ }\bibfield  {title} {\bibinfo {title} {Theoretical study of
  early-time superradiance for atom clouds and arrays},\ }\href
  {https://doi.org/10.1103/PhysRevA.104.063706} {\bibfield  {journal} {\bibinfo
   {journal} {Phys. Rev. A}\ }\textbf {\bibinfo {volume} {104}},\ \bibinfo
  {pages} {063706} (\bibinfo {year} {2021})}\BibitemShut {NoStop}%
\bibitem [{\citenamefont {Fleischhauer}\ and\ \citenamefont
  {Yelin}(1999)}]{Fleishbauer}%
  \BibitemOpen
  \bibfield  {author} {\bibinfo {author} {\bibfnamefont {M.}~\bibnamefont
  {Fleischhauer}}\ and\ \bibinfo {author} {\bibfnamefont {S.~F.}\ \bibnamefont
  {Yelin}},\ }\bibfield  {title} {\bibinfo {title} {Radiative atom-atom
  interactions in optically dense media: Quantum corrections to the
  lorentz-lorenz formula},\ }\href {https://doi.org/10.1103/PhysRevA.59.2427}
  {\bibfield  {journal} {\bibinfo  {journal} {Phys. Rev. A}\ }\textbf {\bibinfo
  {volume} {59}},\ \bibinfo {pages} {2427} (\bibinfo {year}
  {1999})}\BibitemShut {NoStop}%
\bibitem [{\citenamefont {Lin}\ and\ \citenamefont
  {F.Yelin}(2012)}]{AMO_GuindarLin}%
  \BibitemOpen
  \bibfield  {author} {\bibinfo {author} {\bibfnamefont {G.-D.}\ \bibnamefont
  {Lin}}\ and\ \bibinfo {author} {\bibfnamefont {S.}~\bibnamefont {F.Yelin}},\
  }\href@noop {} {\bibfield  {journal} {\bibinfo  {journal} {Adv. At. Mol. Opt.
  Phys.}\ }\textbf {\bibinfo {volume} {61}},\ \bibinfo {pages} {295} (\bibinfo
  {year} {2012})}\BibitemShut {NoStop}%
\bibitem [{\citenamefont {Ma}\ \emph {et~al.}(2022)\citenamefont {Ma},
  \citenamefont {Rubies-Bigorda},\ and\ \citenamefont {Yelin}}]{Hanzhen}%
  \BibitemOpen
  \bibfield  {author} {\bibinfo {author} {\bibfnamefont {H.}~\bibnamefont
  {Ma}}, \bibinfo {author} {\bibfnamefont {O.}~\bibnamefont {Rubies-Bigorda}},\
  and\ \bibinfo {author} {\bibfnamefont {S.~F.}\ \bibnamefont {Yelin}},\ }\href
  {https://arxiv.org/abs/2205.15255} {\bibinfo {title} {Superradiance and
  subradiance in a gas of two-level atoms}} (\bibinfo {year} {2022}),\ \Eprint
  {https://arxiv.org/abs/2205.15255} {arXiv:2205.15255} \BibitemShut {NoStop}%
\bibitem [{Note1()}]{Note1}%
  \BibitemOpen
  \bibinfo {note} {We refer the reader to Ref.~\cite {Hanzhen} for a thorough
  derivation of the full formalism described in the main text.}\BibitemShut
  {Stop}%
\bibitem [{\citenamefont {Keldysh}(1965)}]{keldysh}%
  \BibitemOpen
  \bibfield  {author} {\bibinfo {author} {\bibfnamefont {L.~V.}\ \bibnamefont
  {Keldysh}},\ }\href@noop {} {\bibfield  {journal} {\bibinfo  {journal} {Sov.
  Phys. JETP}\ }\textbf {\bibinfo {volume} {20}},\ \bibinfo {pages} {1018}
  (\bibinfo {year} {1965})}\BibitemShut {NoStop}%
\bibitem [{\citenamefont {Putnam}\ \emph {et~al.}(2016)\citenamefont {Putnam},
  \citenamefont {Lin},\ and\ \citenamefont {Yelin}}]{shifts_small_gray2016}%
  \BibitemOpen
  \bibfield  {author} {\bibinfo {author} {\bibfnamefont {G.}~\bibnamefont
  {Putnam}}, \bibinfo {author} {\bibfnamefont {G.-D.}\ \bibnamefont {Lin}},\
  and\ \bibinfo {author} {\bibfnamefont {S.~F.}\ \bibnamefont {Yelin}},\
  }\href@noop {} {\bibinfo {title} {Collective induced superradiant
  lineshifts}},\ \bibinfo {howpublished} {arXiv:1612.04477} (\bibinfo {year}
  {2016})\BibitemShut {NoStop}%
\bibitem [{\citenamefont {Dyson}(1949)}]{Dyson}%
  \BibitemOpen
  \bibfield  {author} {\bibinfo {author} {\bibfnamefont {F.~J.}\ \bibnamefont
  {Dyson}},\ }\bibfield  {title} {\bibinfo {title} {The $s$ matrix in quantum
  electrodynamics},\ }\href {https://doi.org/10.1103/PhysRev.75.1736}
  {\bibfield  {journal} {\bibinfo  {journal} {Phys. Rev.}\ }\textbf {\bibinfo
  {volume} {75}},\ \bibinfo {pages} {1736} (\bibinfo {year}
  {1949})}\BibitemShut {NoStop}%
\bibitem [{\citenamefont {Akkermans}\ \emph {et~al.}(2008)\citenamefont
  {Akkermans}, \citenamefont {Gero},\ and\ \citenamefont {Kaiser}}]{Akkermans}%
  \BibitemOpen
  \bibfield  {author} {\bibinfo {author} {\bibfnamefont {E.}~\bibnamefont
  {Akkermans}}, \bibinfo {author} {\bibfnamefont {A.}~\bibnamefont {Gero}},\
  and\ \bibinfo {author} {\bibfnamefont {R.}~\bibnamefont {Kaiser}},\
  }\bibfield  {title} {\bibinfo {title} {Photon localization and dicke
  superradiance in atomic gases},\ }\href
  {https://doi.org/10.1103/PhysRevLett.101.103602} {\bibfield  {journal}
  {\bibinfo  {journal} {Phys. Rev. Lett.}\ }\textbf {\bibinfo {volume} {101}},\
  \bibinfo {pages} {103602} (\bibinfo {year} {2008})}\BibitemShut {NoStop}%
\bibitem [{\citenamefont {Müller}\ \emph {et~al.}(2016)\citenamefont
  {Müller}, \citenamefont {Witthaut}, \citenamefont {le~Targat}, \citenamefont
  {Arlt}, \citenamefont {Polzik},\ and\ \citenamefont
  {Hilliard}}]{cigar-shaped}%
  \BibitemOpen
  \bibfield  {author} {\bibinfo {author} {\bibfnamefont {J.~H.}\ \bibnamefont
  {Müller}}, \bibinfo {author} {\bibfnamefont {D.}~\bibnamefont {Witthaut}},
  \bibinfo {author} {\bibfnamefont {R.}~\bibnamefont {le~Targat}}, \bibinfo
  {author} {\bibfnamefont {J.~J.}\ \bibnamefont {Arlt}}, \bibinfo {author}
  {\bibfnamefont {E.~S.}\ \bibnamefont {Polzik}},\ and\ \bibinfo {author}
  {\bibfnamefont {A.~J.}\ \bibnamefont {Hilliard}},\ }\bibfield  {title}
  {\bibinfo {title} {Semi-classical dynamics of superradiant rayleigh
  scattering in a bose–einstein condensate},\ }\href
  {https://doi.org/10.1080/09500340.2016.1207815} {\bibfield  {journal}
  {\bibinfo  {journal} {Journal of Modern Optics}\ }\textbf {\bibinfo {volume}
  {63}},\ \bibinfo {pages} {1886} (\bibinfo {year} {2016})}\BibitemShut
  {NoStop}%
\bibitem [{\citenamefont {Paradis}\ \emph {et~al.}(2008)\citenamefont
  {Paradis}, \citenamefont {Barrett}, \citenamefont {Kumarakrishnan},
  \citenamefont {Zhang},\ and\ \citenamefont {Raithel}}]{Isotropic_spherical}%
  \BibitemOpen
  \bibfield  {author} {\bibinfo {author} {\bibfnamefont {E.}~\bibnamefont
  {Paradis}}, \bibinfo {author} {\bibfnamefont {B.}~\bibnamefont {Barrett}},
  \bibinfo {author} {\bibfnamefont {A.}~\bibnamefont {Kumarakrishnan}},
  \bibinfo {author} {\bibfnamefont {R.}~\bibnamefont {Zhang}},\ and\ \bibinfo
  {author} {\bibfnamefont {G.}~\bibnamefont {Raithel}},\ }\bibfield  {title}
  {\bibinfo {title} {Observation of superfluorescent emissions from
  laser-cooled atoms},\ }\href {https://doi.org/10.1103/PhysRevA.77.043419}
  {\bibfield  {journal} {\bibinfo  {journal} {Phys. Rev. A}\ }\textbf {\bibinfo
  {volume} {77}},\ \bibinfo {pages} {043419} (\bibinfo {year}
  {2008})}\BibitemShut {NoStop}%
\bibitem [{\citenamefont {Stroud}\ \emph {et~al.}(1972)\citenamefont {Stroud},
  \citenamefont {Eberly}, \citenamefont {Lama},\ and\ \citenamefont
  {Mandel}}]{RadiationTrapping_1}%
  \BibitemOpen
  \bibfield  {author} {\bibinfo {author} {\bibfnamefont {C.~R.}\ \bibnamefont
  {Stroud}}, \bibinfo {author} {\bibfnamefont {J.~H.}\ \bibnamefont {Eberly}},
  \bibinfo {author} {\bibfnamefont {W.~L.}\ \bibnamefont {Lama}},\ and\
  \bibinfo {author} {\bibfnamefont {L.}~\bibnamefont {Mandel}},\ }\bibfield
  {title} {\bibinfo {title} {Superradiant effects in systems of two-level
  atoms},\ }\href {https://doi.org/10.1103/PhysRevA.5.1094} {\bibfield
  {journal} {\bibinfo  {journal} {Phys. Rev. A}\ }\textbf {\bibinfo {volume}
  {5}},\ \bibinfo {pages} {1094} (\bibinfo {year} {1972})}\BibitemShut
  {NoStop}%
\bibitem [{\citenamefont {Cummings}\ and\ \citenamefont
  {Dorri}(1983)}]{RadiationTrapping_2}%
  \BibitemOpen
  \bibfield  {author} {\bibinfo {author} {\bibfnamefont {F.~W.}\ \bibnamefont
  {Cummings}}\ and\ \bibinfo {author} {\bibfnamefont {A.}~\bibnamefont
  {Dorri}},\ }\bibfield  {title} {\bibinfo {title} {Exact solution for
  spontaneous emission in the presence of $n$ atoms},\ }\href
  {https://doi.org/10.1103/PhysRevA.28.2282} {\bibfield  {journal} {\bibinfo
  {journal} {Phys. Rev. A}\ }\textbf {\bibinfo {volume} {28}},\ \bibinfo
  {pages} {2282} (\bibinfo {year} {1983})}\BibitemShut {NoStop}%
\bibitem [{\citenamefont {Weiss}\ \emph {et~al.}(2018)\citenamefont {Weiss},
  \citenamefont {Ara{\'{u}}jo}, \citenamefont {Kaiser},\ and\ \citenamefont
  {Guerin}}]{RadiationTrapping_3}%
  \BibitemOpen
  \bibfield  {author} {\bibinfo {author} {\bibfnamefont {P.}~\bibnamefont
  {Weiss}}, \bibinfo {author} {\bibfnamefont {M.~O.}\ \bibnamefont
  {Ara{\'{u}}jo}}, \bibinfo {author} {\bibfnamefont {R.}~\bibnamefont
  {Kaiser}},\ and\ \bibinfo {author} {\bibfnamefont {W.}~\bibnamefont
  {Guerin}},\ }\bibfield  {title} {\bibinfo {title} {Subradiance and radiation
  trapping in cold atoms},\ }\href {https://doi.org/10.1088/1367-2630/aac5d0}
  {\bibfield  {journal} {\bibinfo  {journal} {New Journal of Physics}\ }\textbf
  {\bibinfo {volume} {20}},\ \bibinfo {pages} {063024} (\bibinfo {year}
  {2018})}\BibitemShut {NoStop}%
\bibitem [{\citenamefont {Marti}\ \emph {et~al.}(2018)\citenamefont {Marti},
  \citenamefont {Hutson}, \citenamefont {Goban}, \citenamefont {Campbell},
  \citenamefont {Poli},\ and\ \citenamefont {Ye}}]{3Dlattice_JunYe}%
  \BibitemOpen
  \bibfield  {author} {\bibinfo {author} {\bibfnamefont {G.~E.}\ \bibnamefont
  {Marti}}, \bibinfo {author} {\bibfnamefont {R.~B.}\ \bibnamefont {Hutson}},
  \bibinfo {author} {\bibfnamefont {A.}~\bibnamefont {Goban}}, \bibinfo
  {author} {\bibfnamefont {S.~L.}\ \bibnamefont {Campbell}}, \bibinfo {author}
  {\bibfnamefont {N.}~\bibnamefont {Poli}},\ and\ \bibinfo {author}
  {\bibfnamefont {J.}~\bibnamefont {Ye}},\ }\bibfield  {title} {\bibinfo
  {title} {Imaging optical frequencies with $100\text{ }\text{
  }\ensuremath{\mu}\mathrm{Hz}$ precision and $1.1\text{ }\text{
  }\ensuremath{\mu}\mathrm{m}$ resolution},\ }\href
  {https://doi.org/10.1103/PhysRevLett.120.103201} {\bibfield  {journal}
  {\bibinfo  {journal} {Phys. Rev. Lett.}\ }\textbf {\bibinfo {volume} {120}},\
  \bibinfo {pages} {103201} (\bibinfo {year} {2018})}\BibitemShut {NoStop}%
\bibitem [{\citenamefont {Palacios-Berraquero}\ \emph
  {et~al.}(2017)\citenamefont {Palacios-Berraquero}, \citenamefont {Kara},
  \citenamefont {Montblanch}, \citenamefont {Barbone}, \citenamefont
  {Latawiec}, \citenamefont {Yoon}, \citenamefont {Ott}, \citenamefont
  {Loncar}, \citenamefont {Ferrari},\ and\ \citenamefont
  {Atat{\"u}re}}]{2D_mat_1}%
  \BibitemOpen
  \bibfield  {author} {\bibinfo {author} {\bibfnamefont {C.}~\bibnamefont
  {Palacios-Berraquero}}, \bibinfo {author} {\bibfnamefont {D.~M.}\
  \bibnamefont {Kara}}, \bibinfo {author} {\bibfnamefont {A.~R.-P.}\
  \bibnamefont {Montblanch}}, \bibinfo {author} {\bibfnamefont
  {M.}~\bibnamefont {Barbone}}, \bibinfo {author} {\bibfnamefont
  {P.}~\bibnamefont {Latawiec}}, \bibinfo {author} {\bibfnamefont
  {D.}~\bibnamefont {Yoon}}, \bibinfo {author} {\bibfnamefont {A.~K.}\
  \bibnamefont {Ott}}, \bibinfo {author} {\bibfnamefont {M.}~\bibnamefont
  {Loncar}}, \bibinfo {author} {\bibfnamefont {A.~C.}\ \bibnamefont
  {Ferrari}},\ and\ \bibinfo {author} {\bibfnamefont {M.}~\bibnamefont
  {Atat{\"u}re}},\ }\bibfield  {title} {\bibinfo {title} {Large-scale
  quantum-emitter arrays in atomically thin semiconductors},\ }\href
  {https://doi.org/10.1038/ncomms15093} {\bibfield  {journal} {\bibinfo
  {journal} {Nature Communications}\ }\textbf {\bibinfo {volume} {8}},\
  \bibinfo {pages} {15093} (\bibinfo {year} {2017})}\BibitemShut {NoStop}%
\bibitem [{\citenamefont {Haider}\ \emph {et~al.}(2021)\citenamefont {Haider},
  \citenamefont {Sampathkumar}, \citenamefont {Verhagen}, \citenamefont
  {Nádvorník}, \citenamefont {Sonia}, \citenamefont {Valeš}, \citenamefont
  {Sýkora}, \citenamefont {Kapusta}, \citenamefont {Němec}, \citenamefont
  {Hof}, \citenamefont {Frank}, \citenamefont {Chen}, \citenamefont
  {Vejpravová},\ and\ \citenamefont {Kalbáč}}]{2dmat_2}%
  \BibitemOpen
  \bibfield  {author} {\bibinfo {author} {\bibfnamefont {G.}~\bibnamefont
  {Haider}}, \bibinfo {author} {\bibfnamefont {K.}~\bibnamefont
  {Sampathkumar}}, \bibinfo {author} {\bibfnamefont {T.}~\bibnamefont
  {Verhagen}}, \bibinfo {author} {\bibfnamefont {L.}~\bibnamefont
  {Nádvorník}}, \bibinfo {author} {\bibfnamefont {F.~J.}\ \bibnamefont
  {Sonia}}, \bibinfo {author} {\bibfnamefont {V.}~\bibnamefont {Valeš}},
  \bibinfo {author} {\bibfnamefont {J.}~\bibnamefont {Sýkora}}, \bibinfo
  {author} {\bibfnamefont {P.}~\bibnamefont {Kapusta}}, \bibinfo {author}
  {\bibfnamefont {P.}~\bibnamefont {Němec}}, \bibinfo {author} {\bibfnamefont
  {M.}~\bibnamefont {Hof}}, \bibinfo {author} {\bibfnamefont {O.}~\bibnamefont
  {Frank}}, \bibinfo {author} {\bibfnamefont {Y.-F.}\ \bibnamefont {Chen}},
  \bibinfo {author} {\bibfnamefont {J.}~\bibnamefont {Vejpravová}},\ and\
  \bibinfo {author} {\bibfnamefont {M.}~\bibnamefont {Kalbáč}},\ }\bibfield
  {title} {\bibinfo {title} {Superradiant emission from coherent excitons in
  van der waals heterostructures},\ }\href@noop {} {\bibfield  {journal}
  {\bibinfo  {journal} {Advanced Functional Materials}\ }\textbf {\bibinfo
  {volume} {31}},\ \bibinfo {pages} {2102196} (\bibinfo {year}
  {2021})}\BibitemShut {NoStop}%
\bibitem [{\citenamefont {Feng}\ \emph {et~al.}(2016)\citenamefont {Feng},
  \citenamefont {Wang}, \citenamefont {Zhang}, \citenamefont {Zhang},
  \citenamefont {Lou}, \citenamefont {Zhu},\ and\ \citenamefont
  {Wang}}]{vacancy_1}%
  \BibitemOpen
  \bibfield  {author} {\bibinfo {author} {\bibfnamefont {F.}~\bibnamefont
  {Feng}}, \bibinfo {author} {\bibfnamefont {J.}~\bibnamefont {Wang}}, \bibinfo
  {author} {\bibfnamefont {W.}~\bibnamefont {Zhang}}, \bibinfo {author}
  {\bibfnamefont {J.}~\bibnamefont {Zhang}}, \bibinfo {author} {\bibfnamefont
  {L.}~\bibnamefont {Lou}}, \bibinfo {author} {\bibfnamefont {W.}~\bibnamefont
  {Zhu}},\ and\ \bibinfo {author} {\bibfnamefont {G.}~\bibnamefont {Wang}},\
  }\bibfield  {title} {\bibinfo {title} {Efficient generation of nanoscale
  arrays of nitrogen-vacancy centers with long coherence time in diamond},\
  }\href {https://doi.org/10.1007/s00339-016-0445-5} {\bibfield  {journal}
  {\bibinfo  {journal} {Applied Physics A}\ }\textbf {\bibinfo {volume}
  {122}},\ \bibinfo {pages} {944} (\bibinfo {year} {2016})}\BibitemShut
  {NoStop}%
\bibitem [{\citenamefont {Spinicelli}\ \emph {et~al.}(2011)\citenamefont
  {Spinicelli}, \citenamefont {Dr{\'e}au}, \citenamefont {Rondin},
  \citenamefont {Silva}, \citenamefont {Achard}, \citenamefont {Xavier},
  \citenamefont {Bansropun}, \citenamefont {Debuisschert}, \citenamefont
  {Pezzagna}, \citenamefont {Meijer}, \citenamefont {Jacques},\ and\
  \citenamefont {Roch}}]{vacancy_2}%
  \BibitemOpen
  \bibfield  {author} {\bibinfo {author} {\bibfnamefont {P.}~\bibnamefont
  {Spinicelli}}, \bibinfo {author} {\bibfnamefont {A.}~\bibnamefont
  {Dr{\'e}au}}, \bibinfo {author} {\bibfnamefont {L.}~\bibnamefont {Rondin}},
  \bibinfo {author} {\bibfnamefont {F.}~\bibnamefont {Silva}}, \bibinfo
  {author} {\bibfnamefont {J.}~\bibnamefont {Achard}}, \bibinfo {author}
  {\bibfnamefont {S.}~\bibnamefont {Xavier}}, \bibinfo {author} {\bibfnamefont
  {S.}~\bibnamefont {Bansropun}}, \bibinfo {author} {\bibfnamefont
  {T.}~\bibnamefont {Debuisschert}}, \bibinfo {author} {\bibfnamefont
  {S.}~\bibnamefont {Pezzagna}}, \bibinfo {author} {\bibfnamefont
  {J.}~\bibnamefont {Meijer}}, \bibinfo {author} {\bibfnamefont
  {V.}~\bibnamefont {Jacques}},\ and\ \bibinfo {author} {\bibfnamefont {J.-F.}\
  \bibnamefont {Roch}},\ }\bibfield  {title} {\bibinfo {title} {Engineered
  arrays of nitrogen-vacancy color centers in diamond based on implantation of
  cn-molecules through nanoapertures},\ }\href@noop {} {\bibfield  {journal}
  {\bibinfo  {journal} {New Journal of Physics}\ }\textbf {\bibinfo {volume}
  {13}},\ \bibinfo {pages} {025014} (\bibinfo {year} {2011})}\BibitemShut
  {NoStop}%
\bibitem [{\citenamefont {Sipahigil}\ \emph {et~al.}(2016)\citenamefont
  {Sipahigil}, \citenamefont {Evans}, \citenamefont {Sukachev}, \citenamefont
  {Burek}, \citenamefont {Borregaard}, \citenamefont {Bhaskar}, \citenamefont
  {Nguyen}, \citenamefont {Pacheco}, \citenamefont {Atikian}, \citenamefont
  {Meuwly}, \citenamefont {Camacho}, \citenamefont {Jelezko}, \citenamefont
  {Bielejec}, \citenamefont {Park}, \citenamefont {Lončar},\ and\
  \citenamefont {Lukin}}]{vacancy_3}%
  \BibitemOpen
  \bibfield  {author} {\bibinfo {author} {\bibfnamefont {A.}~\bibnamefont
  {Sipahigil}}, \bibinfo {author} {\bibfnamefont {R.~E.}\ \bibnamefont
  {Evans}}, \bibinfo {author} {\bibfnamefont {D.~D.}\ \bibnamefont {Sukachev}},
  \bibinfo {author} {\bibfnamefont {M.~J.}\ \bibnamefont {Burek}}, \bibinfo
  {author} {\bibfnamefont {J.}~\bibnamefont {Borregaard}}, \bibinfo {author}
  {\bibfnamefont {M.~K.}\ \bibnamefont {Bhaskar}}, \bibinfo {author}
  {\bibfnamefont {C.~T.}\ \bibnamefont {Nguyen}}, \bibinfo {author}
  {\bibfnamefont {J.~L.}\ \bibnamefont {Pacheco}}, \bibinfo {author}
  {\bibfnamefont {H.~A.}\ \bibnamefont {Atikian}}, \bibinfo {author}
  {\bibfnamefont {C.}~\bibnamefont {Meuwly}}, \bibinfo {author} {\bibfnamefont
  {R.~M.}\ \bibnamefont {Camacho}}, \bibinfo {author} {\bibfnamefont
  {F.}~\bibnamefont {Jelezko}}, \bibinfo {author} {\bibfnamefont
  {E.}~\bibnamefont {Bielejec}}, \bibinfo {author} {\bibfnamefont
  {H.}~\bibnamefont {Park}}, \bibinfo {author} {\bibfnamefont {M.}~\bibnamefont
  {Lončar}},\ and\ \bibinfo {author} {\bibfnamefont {M.~D.}\ \bibnamefont
  {Lukin}},\ }\bibfield  {title} {\bibinfo {title} {An integrated diamond
  nanophotonics platform for quantum-optical networks},\ }\href
  {https://doi.org/10.1126/science.aah6875} {\bibfield  {journal} {\bibinfo
  {journal} {Science}\ }\textbf {\bibinfo {volume} {354}},\ \bibinfo {pages}
  {847} (\bibinfo {year} {2016})}\BibitemShut {NoStop}%
\bibitem [{\citenamefont {Zhang}\ and\ \citenamefont
  {M\o{}lmer}(2022)}]{multi_exc_molmer}%
  \BibitemOpen
  \bibfield  {author} {\bibinfo {author} {\bibfnamefont {Y.-X.}\ \bibnamefont
  {Zhang}}\ and\ \bibinfo {author} {\bibfnamefont {K.}~\bibnamefont
  {M\o{}lmer}},\ }\bibfield  {title} {\bibinfo {title} {Free-fermion multiply
  excited eigenstates and their experimental signatures in 1d arrays of
  two-level atoms},\ }\href {https://doi.org/10.1103/PhysRevLett.128.093602}
  {\bibfield  {journal} {\bibinfo  {journal} {Phys. Rev. Lett.}\ }\textbf
  {\bibinfo {volume} {128}},\ \bibinfo {pages} {093602} (\bibinfo {year}
  {2022})}\BibitemShut {NoStop}%
\bibitem [{\citenamefont {Masson}\ and\ \citenamefont
  {Asenjo-Garcia}(2020)}]{2_exc_Ana}%
  \BibitemOpen
  \bibfield  {author} {\bibinfo {author} {\bibfnamefont {S.~J.}\ \bibnamefont
  {Masson}}\ and\ \bibinfo {author} {\bibfnamefont {A.}~\bibnamefont
  {Asenjo-Garcia}},\ }\bibfield  {title} {\bibinfo {title} {Atomic-waveguide
  quantum electrodynamics},\ }\href
  {https://doi.org/10.1103/PhysRevResearch.2.043213} {\bibfield  {journal}
  {\bibinfo  {journal} {Phys. Rev. Research}\ }\textbf {\bibinfo {volume}
  {2}},\ \bibinfo {pages} {043213} (\bibinfo {year} {2020})}\BibitemShut
  {NoStop}%
\bibitem [{\citenamefont {Chang}\ \emph {et~al.}(2018)\citenamefont {Chang},
  \citenamefont {Douglas}, \citenamefont {Gonz\'alez-Tudela}, \citenamefont
  {Hung},\ and\ \citenamefont {Kimble}}]{OtherReservoirs}%
  \BibitemOpen
  \bibfield  {author} {\bibinfo {author} {\bibfnamefont {D.~E.}\ \bibnamefont
  {Chang}}, \bibinfo {author} {\bibfnamefont {J.~S.}\ \bibnamefont {Douglas}},
  \bibinfo {author} {\bibfnamefont {A.}~\bibnamefont {Gonz\'alez-Tudela}},
  \bibinfo {author} {\bibfnamefont {C.-L.}\ \bibnamefont {Hung}},\ and\
  \bibinfo {author} {\bibfnamefont {H.~J.}\ \bibnamefont {Kimble}},\ }\bibfield
   {title} {\bibinfo {title} {Colloquium: Quantum matter built from nanoscopic
  lattices of atoms and photons},\ }\href
  {https://doi.org/10.1103/RevModPhys.90.031002} {\bibfield  {journal}
  {\bibinfo  {journal} {Rev. Mod. Phys.}\ }\textbf {\bibinfo {volume} {90}},\
  \bibinfo {pages} {031002} (\bibinfo {year} {2018})}\BibitemShut {NoStop}%
\end{thebibliography}%

\end{document}